\documentclass[aps,pra,showpacs,citeautoscript,amsmath,amssymb,floatfix,superscriptaddress,bbm, verbatim,amsfonts,changes,10pt,notitlepage,nofootinbib]{revtex4-1}
\pdfoutput=1
\usepackage{amsmath,bbold}
\usepackage{amssymb}
\usepackage{amsmath,latexsym}
\usepackage{epsfig}
\usepackage{color}
\usepackage{amsthm}
\usepackage{hyperref}
\usepackage{appendix}
\usepackage{array}
\usepackage{graphicx}
\usepackage{cancel}
\usepackage{tabularray}
\usepackage{stmaryrd}
\usepackage{subcaption}
\usepackage{tabularx}
\usepackage{soul}
\usepackage{ulem}
\usepackage{mathtools}
\usepackage{feynmf}
\usepackage{setspace}
\usepackage{enumitem}
\usepackage{amssymb}
\usepackage{tikz-cd}
\usepackage{todonotes}
\usepackage{ragged2e}
\newcolumntype{C}{>{\centering\arraybackslash}X}
\newcommand{\lin}{L}

\newcommand{\rind}{+}
\newcommand{\lind}{-}
\newcommand{\q}{q}
\newcommand{\p}{p}
\newcommand{\xq}{\phi}
\newcommand{\pq}{\pi}
\newcommand{\xql}{\phi^-}
\newcommand{\xqr}{\phi^+}
\newcommand{\pql}{\pi^-}
\newcommand{\pqr}{\pi^+}

\def\<{\langle}
\def\>{\rangle}

\newcommand{\cqstate}{\varrho}

\newcommand{\be}{\begin{eqnarray} \begin{aligned}}
\newcommand{\ee}{\end{aligned} \end{eqnarray} }
\newcommand{\benn}{\begin{eqnarray*} \begin{aligned}}
\newcommand{\eenn}{\end{aligned} \end{eqnarray*} }

\newcommand{\ben}{\begin{eqnarray} \begin{aligned}}
\newcommand{\een}{\end{aligned} \end{eqnarray} }

\newcommand{\bc}{\begin{center}}
\newcommand{\ec}{\end{center}}

\newcommand{\Tr}{\mathop{\mathsf{Tr}}\nolimits}

\newcommand{\beq}{\begin{eqnarray} \begin{aligned}}
\newcommand{\eeq}{\end{aligned} \end{eqnarray} }
\newcommand{\bea}{\begin{array}}
\newcommand{\eea}{\end{array}}

\newcommand{\bee}{\begin{enumerate}}
\newcommand{\eee}{\end{enumerate}}
\newcommand{\bei}{\begin{itemize}}
\newcommand{\eei}{\end{itemize}}

\usepackage{amsfonts}

\def\01{\{0,1\}}

\newcommand{\cE}{\mathcal{E}}

\newcommand{\str}{{\vec{x}}}

\def\<{\langle}
\def\>{\rangle}

\def\s{\,\,\,\,}

\newtheorem*{rep@theorem}{\rep@title}
\newcommand{\newreptheorem}[2]{%
\newenvironment{rep#1}[1]{%
 \def\rep@title{#2 \ref{##1} (restatement)}%
 \begin{rep@theorem}}%
 {\end{rep@theorem}}}
\makeatother

\newreptheorem{thm}{Theorem}
\newreptheorem{lem}{Lemma}

\def\T{\bf T}

\def\z{{z}}
\def\u{{ u}}

\def\g{{g}}

\def\T00{{\bf T_{NN}}}

\def\0mom{{\bar{\Gamma}^{\alpha\beta}(\z)}}
\def\1mom{{\Gamma^{\alpha\beta}_1(\z)}}
\def\2mom{{\Gamma^{\alpha\beta}_2(\z)}}

\tikzset{
my loop/.style={to path={
.. controls +(50:1.1) and +(130:1.1) .. (\tikztotarget) \tikztonodes}}}
\tikzset{
my loop2/.style={to path={
.. controls +(50:0.8) and +(130:0.8) .. (\tikztotarget) \tikztonodes}}}
\tikzset{
my loop3/.style={to path={
.. controls +(10:0.8) and +(95:0.8) .. (\tikztotarget) \tikztonodes}}}
\tikzset{
my loop4/.style={to path={
.. controls +(-50:0.5) and +(10:0.4) .. (\tikztotarget) \tikztonodes}}}
\tikzset{
my loop5/.style={to path={
.. controls +(230:0.5) and +(170:0.4) .. (\tikztotarget) \tikztonodes}}}

\long\def\diagramone{\begin{tikzpicture}
\draw[fill=black, dashed](0,0)circle(2pt)  node[anchor=north east]{$\phi^+$}   -- +(2cm,0) circle(2pt)node[anchor=north west]{$\phi^+$};
\path (0.9cm,-0.2cm) node[anchor=north]{ $-\frac{i\hbar}{m_{\phi}^2}$} (0cm,+0.1cm) ;
\end{tikzpicture}\hspace{0.5cm}
\begin{tikzpicture}
\draw[fill=black, dotted](0,0)circle(2pt)  node[anchor=north east]{$\phi^-$}   -- +(2cm,0) circle(2pt)node[anchor=north west]{$\phi^-$};
\path (1.1cm,-0.2cm) node[anchor=north]{ $\frac{i\hbar}{m_{\phi}^2}$} (0cm,+0.1cm) ;
\end{tikzpicture}\hspace{0.5cm}
\begin{tikzpicture}
\draw[fill=black](0,0)circle(2pt)  node[anchor=north east]{$q$}   -- +(2cm,0) circle(2pt)node[anchor=north west]{$q$};
\path (1.1cm,-0.2cm) node[anchor=north]{ $\frac{D_2}{m_q^4}$} (0cm,+0.1cm) ;
\end{tikzpicture} 
}
\long\def\interactiondiagram{\begin{tikzpicture}
\coordinate (A) at (1,0);
\node[xshift=0.8cm] at (A) {$-\frac{\lambda m_q^2}{2D_2}$};
\draw[fill=black, dashed](0,0)   -- +(1cm,0) circle(2pt)node[anchor=north west]{} -- (1.5cm,0.865cm);
\draw[fill=black](1cm,0) -- (1.5cm,-0.865cm);
\end{tikzpicture}\hspace{0.5cm}
\begin{tikzpicture}
\coordinate (A) at (1,0);
\node[xshift=0.8cm] at (A) {$-\frac{\lambda m_q^2}{2D_2}$};
\draw[fill=black, dotted](0,0)   -- +(1cm,0) circle(2pt)node[anchor=north west]{} -- (1.5cm,0.865cm);
\draw[fill=black](1cm,0) -- (1.5cm,-0.865cm);
\end{tikzpicture}}

\long\def\interactiondiagramsix{\begin{tikzpicture}
\coordinate (A) at (2,0);
\node[xshift=0.8cm] at (A) {$-\frac{12 \lambda^2 }{D_2}$};
\draw[fill=black, dotted](0cm,0)   -- + (1cm,0) circle(2pt)node[anchor=north west]{};
\draw[fill=black](1cm,0) -- (0.29cm,0.71cm);
\draw[fill=black,dotted](1cm,0) -- (1.71cm,-0.71cm);
\draw[fill=black](1cm,0) -- (1.71cm,0.71cm);
\draw[fill=black,dotted](1cm,0) -- (0.29cm,-0.71cm);
\draw[fill=black, dotted](1cm,0) -- (2cm,0cm);
\end{tikzpicture}\hspace{0.5cm}
\begin{tikzpicture}
\coordinate (A) at (2,0);
\node[xshift=0.8cm] at (A) {$-\frac{12 \lambda^2 }{D_2}$};
\draw[fill=black, dotted](0cm,0)   -- + (1cm,0) circle(2pt)node[anchor=north west]{};
\draw[fill=black](1cm,0) -- (0.29cm,0.71cm);
\draw[fill=black,dotted](1cm,0) -- (1.71cm,-0.71cm);
\draw[fill=black,dotted](1cm,0) -- (1.71cm,0.71cm);
\draw[fill=black,dotted](1cm,0) -- (0.29cm,-0.71cm);
\draw[fill=black](1cm,0) -- (2cm,0cm);
\end{tikzpicture}\hspace{0.5cm}
\begin{tikzpicture}
\coordinate (A) at (2,0);
\node[xshift=0.8cm] at (A) {$-\frac{12 \lambda^2 }{D_2}$};
\draw[fill=black, dotted](0cm,0)   -- + (1cm,0) circle(2pt)node[anchor=north west]{};
\draw[fill=black](1cm,0) -- (0.29cm,0.71cm);
\draw[fill=black](1cm,0) -- (1.71cm,-0.71cm);
\draw[fill=black,dotted](1cm,0) -- (1.71cm,0.71cm);
\draw[fill=black,dotted](1cm,0) -- (0.29cm,-0.71cm);
\draw[fill=black,dotted](1cm,0) -- (2cm,0cm);
\end{tikzpicture}
}
\long\def\interactiondiagramsixdash{\begin{tikzpicture}
\coordinate (A) at (2,0);
\node[xshift=0.8cm] at (A) {$-\frac{12 \lambda^2 }{D_2}$};
\draw[fill=black, dashed](0cm,0)   -- + (1cm,0) circle(2pt)node[anchor=north west]{};
\draw[fill=black](1cm,0) -- (0.29cm,0.71cm);
\draw[fill=black,dashed](1cm,0) -- (1.71cm,-0.71cm);
\draw[fill=black](1cm,0) -- (1.71cm,0.71cm);
\draw[fill=black,dashed](1cm,0) -- (0.29cm,-0.71cm);
\draw[fill=black, dashed](1cm,0) -- (2cm,0cm);
\end{tikzpicture}\hspace{0.5cm}
\begin{tikzpicture}
\coordinate (A) at (2,0);
\node[xshift=0.8cm] at (A) {$-\frac{12 \lambda^2 }{D_2}$};
\draw[fill=black, dashed](0cm,0)   -- + (1cm,0) circle(2pt)node[anchor=north west]{};
\draw[fill=black](1cm,0) -- (0.29cm,0.71cm);
\draw[fill=black,dashed](1cm,0) -- (1.71cm,-0.71cm);
\draw[fill=black,dashed](1cm,0) -- (1.71cm,0.71cm);
\draw[fill=black,dashed](1cm,0) -- (0.29cm,-0.71cm);
\draw[fill=black](1cm,0) -- (2cm,0cm);
\end{tikzpicture}\hspace{0.5cm}
\begin{tikzpicture}
\coordinate (A) at (2,0);
\node[xshift=0.8cm] at (A) {$-\frac{12 \lambda^2 }{D_2}$};
\draw[fill=black, dashed](0cm,0)   -- + (1cm,0) circle(2pt)node[anchor=north west]{};
\draw[fill=black](1cm,0) -- (0.29cm,0.71cm);
\draw[fill=black](1cm,0) -- (1.71cm,-0.71cm);
\draw[fill=black,dashed](1cm,0) -- (1.71cm,0.71cm);
\draw[fill=black,dashed](1cm,0) -- (0.29cm,-0.71cm);
\draw[fill=black,dashed](1cm,0) -- (2cm,0cm);
\end{tikzpicture}
}
\begin{document}

\title{Path integrals for classical-quantum dynamics }

\author{Jonathan Oppenheim}

%%\email{j.oppenheim@ucl.ac.uk}
\affiliation{Department of Physics and Astronomy, University College London, Gower Street, London WC1E 6BT, United Kingdom}
\author{Zachary Weller-Davies}

\affiliation{Perimeter Institute for Theoretical Physics, Waterloo, Ontario, Canada}
\affiliation{Department of Physics and Astronomy, University College London, Gower Street, London WC1E 6BT, United Kingdom}

\begin{abstract}
Consistent dynamics which couples classical and quantum degrees of freedom exists. This dynamics is linear in the hybrid state, completely positive and trace preserving. Starting from completely positive classical-quantum master equations, we derive a general path integral representation for such dynamics in terms of a classical-quantum action, which includes the necessary and sufficient conditions for complete positivity and trace preservation. The path integral we study is a generalization of the Feynman path integral for quantum systems, and the stochastic path integral used to study classical stochastic processes, allowing for interaction between the classical and quantum systems. When the classical-quantum Hamiltonian is at most quadratic in the momenta we are able to derive a configuration space path integral, providing a map between master equations and covariant classical-quantum path integrals.
\end{abstract}

\maketitle

\section{introduction}\label{sec: Intro}

Recently, there has been progress in understanding the dynamics of classical-quantum systems, where one system can be treated as classical and the other quantum mechanically. Examples of consistent classical-quantum (CQ) master equations, originally introduced in \cite{blanchard1995event,diosi1995quantum}, have since been studied in a variety of different contexts \cite{alicki2003completely, poulinKITP2, Oppenheim:2018igd,oppenheim2020objective}, including gravity \cite{Diosi:2011vu, Oppenheim:2018igd,pqconstraints,dec_Vs_diff}, and can be shown to be completely positive, trace preserving (CPTP), and preserve the split between classical and quantum degrees of freedom \cite{blanchard1995event,Diosi:1995qs, alickiCP, oppenheim2020objective}. The CPTP condition is required for the dynamics to respect positivity and normalisation of probabilities \cite{alicki2007quantum}. Other approaches to classical-quantum dynamics include sourcing classical degrees of freedom via feedback and measurement of quantum matter \cite{Kafri_2014, Kafri_2015, TilloySemiClassical, Diosi2014}, which also lead to completely positive classical-quantum master equations. The most general form of CPTP classical-quantum dynamics has been derived \cite{Oppenheim:2018igd,UCLPawula}, and takes an analogous form to that of the GSKL or Lindblad equation \cite{GSKL,Lindblad:1975ef} in open quantum systems and the rate equation in classical dynamics. 

In the master equation picture, the complete positivity, and general consistency, of the dynamics is manifest. However, in a variety of contexts a path integral approach is perhaps more useful. For example, some numerical simulations are better suited to path integral methods \cite{pathIntSim,monteCarloPI, StochasticPiaction, StochasticPIcont1, PhysRevD.33.1643,PhysRevB.78.045308,PhysRevB.78.045308,PhysRevA.56.2334}, especially when saddle point approximations are valid. For practical applications, it is useful to note that classical-quantum dynamics can be viewed as the natural framework to discuss quantum theory when measurements are involved, which is particularly relevant for quantum control procedures. Indeed, the most general operation one is allowed to perform in standard quantum theory is described by a series of CPTP maps which are performed conditioned on the outcomes of measurements, and is described by the classical-quantum map in Equation \eqref{eq: CPmap}. On the other hand, CQ dynamics is the framework to consider theories with a classical field, whether fundamental or effective, and a path integral approach allows one to impose space-time and gauge symmetries, as well as the possibility to enforce the modern principles used when studying effective field theories \cite{GeorgiEFT}.

In quantum mechanics, it is well known that one can derive the path integral approach from the Schr{\" o}dinger equation, and also from the more general Lindblad equation arising from open quantum systems \cite{ FeynmanVernon, Sieberer_2016}. It is perhaps less well known that one can do the same for classical dynamics, arriving at an equivalence between general master equations and path integrals 
\cite{Onsager1953Fluctuations,freidlin1998random,Weber_2017, Kleinert}. For example, a Brownian particle whose conditional probability distribution $P(x,p|x',p')$ evolves according to the Fokker-Plank equation has an equivalent description in terms of a path integral which is (up to factors of $i, \hbar$) the same as the standard path integral of quantum mechanics.

In this work we shall make use of the recent developments in the understanding of classical-quantum master equations to write down a classical-quantum path integral, equivalent to dynamics which is CPTP. Specifically, using the most general form of CPTP classical-quantum dynamics introduced in \cite{Oppenheim:2018igd,UCLPawula}, we associate a path integral to \textit{any} CQ master equation, from which the conditions on complete positivity can easily be read off. The general result is given by Equation \eqref{eq: generalAction}. We also study CP classical-quantum path integrals without resorting to master equation methods in an accompanying paper \cite{UCLPIShort}. In Table \ref{tab: MasterEquationTable} we compare the standard Feynman path integral for quantum systems, the classical path integral for stochastic systems, and the classical-quantum hybrid path integrals we construct in this work.

\begin{table}
\label{tab:pathintegrals}
\caption{A table representing the classical, quantum, and classical-quantum path integrals.}
\small
\begin{subtable}{\linewidth}
\begin{tblr}{
  colspec = {X[2cm,c]X[c]},
  stretch = 0,
  rowsep = 6pt,
  hlines = {black, 0.5pt},
  vlines = {black, 0.5pt},
}
 & Classical stochastic  \\

    \parbox{\linewidth}{Path integral} & $p(q, p,t_f) = \int \mathcal{D} q \mathcal{D} p \ e^{iS_C[q,p]} \delta( \dot{q} - \frac{\partial H}{\partial p}) p(q,p,t_i) $  \\ 

\parbox{\linewidth}{Action} & $iS_C =-\int_{t_i}^{t_f} dt \frac{1}{2} \ (\frac{\partial H}{\partial q} + \dot{p}) D_2^{-1} (\frac{\partial H}{\partial q}  + \dot{p}) $  \\ 

\parbox{\linewidth}{CP condition} &     $D^{-1}_2$ a positive (semi-definite) matrix, $D^{-1}_2 \succeq 0 $  \\

\end{tblr}

\caption{The path integral for continuous, stochastic classical dynamics \cite{Onsager1953Fluctuations, freidlin1998random,Weber_2017, Kleinert} in phase space generated by a stochastic Hamiltonian. One sums over all classical configurations $(q,p)$ with a weighting according to the difference between the classical path and its expected force $-\frac{\partial H}{\partial q}$, by an amount characterized by the diffusion matrix $D_{2}$. In the case where the force is determined by a Lagrangian $L_C$, the action $S_C$ describes a suppression of paths away from the Euler-Lagrange equations $iS_C =-\int_{t_i}^{t_f} dt \frac{1}{2} (\frac{\delta L_C }{\delta q_i}) (D_2^{-1})^{ij} (\frac{\delta L_C }{\delta q_j}) $, by an amount determined by the diffusion coefficient $D_2$. The most general form of classical path integral is given by Equation \eqref{eq: genclassicalPI}.  } 
\end{subtable}
\vspace{0.1cm}

\begin{subtable}{\linewidth}
\begin{tblr}{
  colspec = {X[2cm,c]X[c]},
  stretch = 0,
  rowsep = 6pt,
  hlines = {black, 0.5pt},
  vlines = {black, 0.5pt},
}
 & Quantum    \\

    \parbox{\linewidth}{Path integral} & $\rho(\phi^{\pm}, t_f) = \int \mathcal{D} \phi^{\pm} \ e^{iS[\phi^+] - iS[\phi^- ] + iS_{FV}[\phi^+, \phi^-] } \rho(\phi^{\pm}, t_i)$   \\ 

\parbox{\linewidth}{Action} & $\displaystyle 
\begin{aligned}
& S[\xq]  = \int_{t_i}^{t_f} dt \big( \frac{1}{2}\dot{\xq}^2 + V(\xq) \big), \\
 & iS_{FV} =  \int_{t_i}^{t_f} dt \big ( D^{\alpha \beta}_0  \lin_{\alpha}^{\rind} \lin_{\beta}^{* \lind} - \frac{1}{2} D^{\alpha \beta}_0 (L^{* \lind}_{\beta}\lin_{\alpha}^{\lind} + L^{*\rind}_{\beta}\lin_{\alpha}^{\rind} ) \big) 
 \end{aligned}$ \\

\parbox{\linewidth}{CP condition} &   $D_0^{\alpha \beta}$ a positive (semi-definite) matrix, $D_0 \succeq 0$. \\

\end{tblr}

\caption{\label{tab: quantumPI}The path integral for a general autonomous quantum system, here taken to be $\phi$. The quantum path integral is doubled since it includes a path integral over both the bra and ket components of the density matrix, here represented using the $\pm$ notation. In the absence of the Feynman Vernon term $S_{FV}$ \cite{FeynmanVernon}, the path integral represents a quantum system evolving unitarily with an action $S[\phi]$. When the Feynman Vernon action $S_{FV}$ is included, the path integral describes the path integral for dynamics undergoing Lindbladian evolution \cite{Lindblad:1975ef, GSKL} with Lindblad operators $L_\alpha(\phi)$. Because of the $\pm$ cross terms, the path integral no-longer preserves the purity of the quantum state and there will generally be decoherence  by an amount determined by $D_0$. By way of example, taking $D^{\alpha \beta}_0=D_0$ and $\lin_{\alpha}^{\pm}=\phi^\pm (x)$ a local field, results in a Feynman-Vernon term
$iS_{FV} =  - D_0 \int_{t_i}^{t_f} dt dx \big (\phi^\lind(x)-\phi^\rind(x)\big) ^2$ which decoheres the state in the $\phi(x)$ basis: off-diagonal terms in the density matrix where $\phi^\rind(x)$ is different to $\phi^\lind(x)$ are suppressed.
}
\end{subtable}

\vspace{0.3cm}

\begin{subtable}{\linewidth}
\begin{tblr}{
  colspec = {X[2cm,c]X[c]},
  stretch = 0,
  rowsep = 6pt,
  hlines = {black, 0.5pt},
  vlines = {black, 0.5pt},
}
 & Classical-quantum   \\

    \parbox{\linewidth}{Path integral} & $ \rho(q,p, \phi^{\pm}, t_f) = \int \mathcal{D} q  \mathcal{D} p \mathcal{D} \phi^{\pm} \ e^{iS_C[q,p] + iS[\phi^+] - iS[\phi^{-} ] + iS_{FV}[\phi^{\pm}]+ iS_{CQ}[q,p, \phi^{\pm}]  } \delta( \dot{q} - \frac{p}{m}) \rho(q,p,\phi^{\pm}, t_i)
    $\\ 

\parbox{\linewidth}{Action} & $\displaystyle 
\begin{aligned}
& iS_C[z] + iS_{CQ}[z,\phi^{\pm}] = -\frac{1}{2} \int_{t_i}^{t_f} dt \ D_2^{-1}\big( \frac{\partial H_C}{\partial q}+ \frac{1}{2} \frac{\partial V_I[q,\phi^+]}{\partial q } + \frac{1}{2} \frac{\partial V_I[q,\phi^-]}{\partial q } + \dot{p}\big)^2.
 \end{aligned}$ \\

\parbox{\linewidth}{CP condition} &   $D_0 \succeq 0, D_2 \succeq 0$ and $4D_2 \succeq  D_0^{-1}$ \\

\end{tblr}
\caption{The phase space path integral for continuous, autonomous classical-quantum dynamics. The path integral is a sum over all classical paths of the variables $q,p$, as well as a sum over the doubled quantum degrees of freedom $\phi^{\pm}$. The action contains the purely quantum term from the quantum path integral in Table \ref{tab: quantumPI}, but also includes the term $iS_C+iS_{CQ}$. This suppresses paths away from the drift, which is sourced by both purely classical terms described by the Hamiltonian $H_C$ and the back-reaction of the quantum systems on the classical ones, described by a classical-quantum interaction potential $V_I$. The action acts to suppress paths which deviate from the $\pm$ averaged Hamilton's equations (see Equation \eqref{eq: hamDrift}). The most general form of classical-quantum path integral is given by Equation \eqref{eq: generalAction}. Under certain conditions, namely when the classical-quantum action 
 \eqref{eq: generalAction} is quadratic in momenta, one can arrive at a configuration space path integral, where paths deviating from the $\pm$ averaged Euler-Lagrange equations as suppressed (see Equation \eqref{eq: continuousActionPositionSpace}). In order for the dynamics to be completely positive, the decoherence-diffusion trade-off $4D_2 \succeq D_0^{-1}$ must be satisfied \cite{UCLPawula,dec_Vs_diff}, where $D_0^{-1}$ is the generalized inverse of $D_0$, which must be positive semi-definite.. As a consequence, there must be both a Feynman-Vernon term $D_0$, and deviation from paths away from their expected drift due to the diffusion coefficient $D_2$, and both effects cannot be made small. When the trade-off is saturated, the path integral preserves purity of the quantum state, conditioned on the classical degree of freedom\cite{UCLHealing} (see Section \ref{sec: saturate}). 
}
\end{subtable}
\label{tab: MasterEquationTable} 
\end{table}

Classical-quantum path integrals have appeared previously \cite{MakriReview,LambertMakri,LambertMakri2,MakriExample,mccaul2020win}. These may be valid when applied to some initial probability densities, but generally lead to negative probabilities since the dynamics is not completely positive on all initial states. Here, we consider dynamics which is CPTP on all states at all times. Of particular relevance is the class of continuous master equations \cite{Diosi:1995qs,Diosi:2011vu}, the most general form of which was introduced in \cite{UCLPawula}. We find these path integrals have a natural decomposition into a pure classical part, representing the stochastic nature of the classical degrees of freedom, a pure quantum part, which includes a Feynman-Vernon term, and a classical-quantum part -- which acts to exponentially suppress the paths which deviate from the averaged equations of motion -- as summarized by Table \ref{tab: MasterEquationTable}.
Under certain conditions, namely when the dynamics is at most quadratic in momenta, we can integrate out the momenta to arrive at a configuration space path integral. In the case where the dynamics is approximately Hamiltonian, as in \cite{Diosi:2011vu,Oppenheim:2018igd, oppenheim2020objective}, the configuration space path integral acts to exponentially suppress the paths which deviate from paths solving the averaged Euler-Lagrange equations. 

The final form we find motivates a general form of configuration space path integrals, summarized by Equation \eqref{eq: fieldConfigurationSpace2}. In \cite{UCLPIShort} we prove such path integrals are completely positive without resorting to master equation methods, meaning the general form is valid even when higher derivative terms are included. As a result, these path integrals provide a general framework to construct classical-quantum theories which respect space-time symmetries. In this work we are primarily interested in deriving CQ path integrals from master equations, but for completeness we also include a discussion of their more general form. These are discussed in more detail in \cite{UCLPIShort} where they are used to construct CQ path integrals for gravity including a diffeomorphism invariant theory based on the trace of Einstein's equations. These CQ path integrals can be thought of as an effective theory where space-time is treated as classical. On the other hand, if taken as fundamental, the parameter space of the theory can be experimentally constrained via the decoherence diffusion trade-off \cite{dec_Vs_diff}, which has already been used to constrain theories with a fundamentally classical gravitational field.
We will find that the trade-off plays a special role here. When it is saturated, the path integral takes on a particularly simple form.

The outline of the paper is as follows. In section \ref{sec: cqdnamics} we briefly review the CQ master equation, where we discuss the necessary and sufficient conditions for complete positivity, as well as reviewing the classical-quantum Kramers-Moyal expansion \cite{Oppenheim:2018igd} -- a tool which is helpful in deriving the path integral. In section \ref{sec: derivationGeneral}, we show how one can perform a Trotterization \cite{trotter} of the CQ master equation to arrive at a general CQ path integral in terms of the quantum variables, the classical variables and ``response variables" denoted $\u$, which are often found in path integral approaches to classical dynamics \cite{JansennRNCFT}. In section \ref{sec: derivationContinuous} we show that for the continuous class of master equations \cite{UCLPawula}, it is possible to integrate out the response variables to get a path integral in terms of the classical and quantum degrees of freedom alone. In the case where the Lindblad operators and Hermitian, or when the decoherence-diffusion trade-off is saturated, the path integral takes on a particularly simple form. This is shown in Subsections \ref{sec: hermitian} and \ref{sec: saturate} respectively.
In section \ref{sec: configurationSpace} we study the configuration space path integral. In the case the dynamics is approximately Hamiltonian, we show that the result of the path integral action is to exponentially suppress the paths which deviate from the averaged Euler-Lagrange equations, with all the information about the classical-quantum interaction encoded in a proto-action $W_{CQ}$. In section \ref{sec: cqfieldPI}, we discuss the path integral for interacting classical and quantum fields. We give an example of a CQ master equation which has a Lorentz invariant path integral. For completeness we also review the covariant path integral formalism introduced in our companion paper \cite{UCLPIShort}. We conclude by mentioning possible directions for future research. 

The appendices contain examples of path integrals. In Appendix \ref{sec: continousmeasurement} we derive the path integral for the most general Markovian\footnote{More precisely, we consider dynamics which is {\it autonomous}, meaning that the coupling constants of the theory don't depend on the time. By including an auxillary classical variable which acts as a clock, we can accommodate the case where the coupling constants are time dependent, so long as they satisfy the same positivity conditions as in the time-independent case.} continuous measurement procedure, where one also allows for classical control and feedback. In Appendix \ref{sec: peturbative} we add a source term to the path integral to allow us to compute correlation functions, and discuss a simple toy example to illustrate how one calculate correlation functions using CQ Feynman diagrams. In Appendix \ref{app: covariantfields} we study examples of CQ path integrals which maintain space-time symmetries such as Lorentz and diffeomorphism invariance.
\section{Classical-Quantum dynamics} \label{sec: cqdnamics}
In this section, we briefly review the master equation governing classical-quantum dynamics as well as introduce the tools which are necessary in deriving a CQ path integral. Of particular importance is the Kramers-Moyal expansion \cite{KRAMERS1940284, Moyal} of the dynamics, whose CQ version \cite{Oppenheim:2018igd} is presented as Equation \eqref{eq: expansion}. In classical Markovian dynamics, the Kramers-Moyal expansion is the starting point to obtaining a path integral representation of the dynamics \cite{Weber_2017}; we find the same in the classical-quantum case, with the moments of the classical-quantum map appearing in the exponent of the path integral. The complete positivity of the dynamics translates to positivity conditions on the moments, the consequences of which have been explored in detail in \cite{dec_Vs_diff}.

We assume the classical degrees of freedom are described by a continuous measurable space $\mathcal{M}$, and we will generically denote elements of the space by $\z$. For example, we could take the classical degrees of freedom to be position and momenta in which case $\mathcal{M}= \mathbb{R}^{2}$ and $\z= (q,p)$. The quantum degrees of freedom are described by a Hilbert space $\mathcal{H}$. We discuss the case of fields separately, and less rigorously, in Section \ref{sec: cqfieldPI}. Given the Hilbert space, we denote the set of positive semi-definite operators as $S(\mathcal{H})$. Then the CQ object defining the state of the CQ system at a given time is a map $\cqstate : \mathcal{M} \to  S(\mathcal{H})$ subject to a normalization constraint $\int_{\mathcal{M}} d\z \Tr_{\mathcal{H}}{\cqstate} =1$. To put it differently, we associate to each classical degree of freedom a un-normalized density operator, $\cqstate(\z)$, such that $\Tr_{\mathcal{H}}{\cqstate} = p(\z) \geq 0$ is a normalized probability distribution over the classical degrees of freedom and $\int_{\mathcal{M}} d\z \cqstate(z) $ is a normalized density operator on $\mathcal{H}$. 

It has been shown~\cite{Oppenheim:2018igd} that any dynamics which maps CQ states onto themselves, if taken to be linear and autonomous, will be completely positive if and only if it can be written in the form
 \begin{equation}\label{eq: CPmap}
 \cqstate(\z,t+ \delta t) =  \int dz' \Lambda(\z|\z',\delta t) ( \cqstate(\z',t)) = \int d\z' \Lambda^{\mu\nu}(\z|\z',\delta t) \lin_{\mu}(\z|\z',\delta t)\cqstate(\z',t) \lin_{\nu}^{\dag}(\z|\z',\delta t),
\end{equation}  
where $ \Lambda(\z|\z',\delta t)$ is a completely positive map for each $\z,\z'$\footnote{Since the classical space is continuous, what we technically ask is that the kernel $\hat{\Lambda} (\z)$, defined via $\hat{\Lambda}(\z)(\psi) = \int dz'\Lambda(\z|\z') \psi(\z') $, is completely positive for all $\z$. Picking $\psi(\z) = \delta(\z,\z')$ then implies that $\Lambda(\z|\z')$ must be CP for all $\z,\z'$. }, the $\lin_{\mu}(\z|\z',\delta t)$ are any set of \textit{Lindblad} operators and $\Lambda^{\mu\nu}(\z|\z',\delta t)$ is a positive measure over $\z,\z'$.
The normalization of probabilities requires
\begin{equation}\label{eq: prob}
\int d\z  \Lambda^{\mu\nu}(\z|\z',\delta t) \lin_{\nu}^{\dag}(\z|\z',\delta t)\lin_{\mu}(\z|\z',\delta t) =\mathbb{I}.
\end{equation}

We now perform a moment expansion of the dynamics in a classical-quantum version of the Kramers-Moyal expansion. We begin by introducing the moments of the transition amplitude 
\begin{equation}\label{eq: moments}
M^{\mu \nu}_{n, i_1 \dots i_n} (\z',\delta t) = \int d\z \ \Lambda^{\mu \nu}(\z|\z',\delta t) (z-z')_{i_1} \dots (z-z')_{i_n},
\end{equation}
where the subscripts $i_j \in \{1, \dots d \}$ label the different components of the vectors $(\z-\z')$. For example, in the case where $d=2$ and the classical degrees of freedom are position and momenta of a particle, $\z=(z_1,z_2) = (q,p)$, then we have $(z-z') = (z_1- z'_1, z_2 - z_2') = ( q-q',p-p')$. The components are then given by $(z-z')_1 = (q-q')$ and $(z-z')_2 = (p-p')$.  $M^{\mu \nu}_{n, i_1 \dots i_n} (z',\delta t)$ is seen to be an $n$'th rank tensor with $d^n$ components. It is important to note that the moments are not independent from each other, due to the condition that the dynamics be CP. In particular, via Equation \eqref{eq: moments}, the condition that $\Lambda^{\mu \nu}(z|z')$ defines a completely positive map transfers to positivity conditions on the moments $M^{\mu \nu}_{n, i_1 \dots i_n}$. It will also be useful to introduce the short time expansion of  Equation \eqref{eq: moments}
\begin{equation}
M^{\mu \nu}(\z',\delta t)_{n, i_1 \dots i_n} =  \delta^{\mu}_0 \delta^{\nu}_0 + \delta t  D^{\mu \nu}_{n, i_1 \dots i_n}(\z',t)  + O(\delta t^2),
\label{eq: defM}
\end{equation} 
which implicitly defines the short time moments $D^{\mu \nu}_{n, i_1 \dots i_n}(\z',t)$.
We define the characteristic function, which is the Fourier transform of the transition amplitude
\begin{equation}
C^{\mu \nu}(\u,\z',\delta t) = \int d\z e^{i \u \cdot (\z-\z')} \Lambda^{\mu \nu}(\z|\z', \delta t) = \sum_{n=0}^{\infty} \frac{(i^n) u_{i_1} \dots u_{i_n}}{n!} M^{\mu \nu}_{n, i_1 \dots i_n}(\z',\delta t),
\end{equation}
and we shall frequently refer to the dual variables $\u$ as \textit{response variables}. Taking the inverse Fourier transform, we can relate the transition amplitude to its moments
\begin{equation}\label{eq: shorttimeLambda}
\Lambda^{\mu \nu}(\z|\z',\delta t) =\int d \u \ e^{-i \u \cdot (\z-\z')}  C^{\mu \nu}(\u,\z',\delta t)  = \sum_{n=0}^{\infty}  \frac{M^{\mu \nu}_{n, i_1 \dots i_n}(\z',\delta t)}{n!}   \frac{1}{(2 \pi)^d} \int d \u \ \ e^{-i \u \cdot (\z-\z')} (i^n) u_{i_1} \dots u_{i_n}.
\end{equation} 
Inserting Equation \eqref{eq: shorttimeLambda} into the CQ map \eqref{eq: CPmap}, and making a short time expansion using Equation \eqref{eq: moments}, we can write the state at $t+\delta t$ in terms of the coefficients $D^{\mu \nu}_{n, i_1 \dots i_n}(\z')$ and the state at $t$ as
\begin{equation}\label{eq: transitionAmplitude}
\cqstate(\z,t+ \delta t) =    \frac{1}{(2 \pi)^d} \int du d\z' \  e^{-i u (\z-\z')} \left( \cqstate(z',t)+  \sum_{n=0}^{\infty}  \delta t (i^n) \frac{ u_{i_1} \dots u_{i_n}   }{n!}  D^{\mu \nu}_{n, i_1 \dots i_n}(\z',t) \lin_{\mu}\cqstate(\z',t) \lin_{\nu}^{\dag} \right).
\end{equation}
Equation \eqref{eq: transitionAmplitude} will be a key equation in deriving the CQ path integral. The moments appearing in Equation \eqref{eq: transitionAmplitude} can be related to physical quantities. For example, the first and second moments $D_1,D_2$ of the probability transition amplitude characterize the amount of drift and diffusion in the system, whilst the zeroth moment $D_0$ can be related to the amount of decoherence of the quantum system. One needs to remember that the moments are not independent of each other and using the conservation of probability one can eliminate $D_{0}^{00}$ in favour of the other coefficients \cite{Oppenheim:2018igd}. By taking the limit $\delta t \to 0$, and using the probability preserving condition in \eqref{eq: prob}, one can write a CQ master equation of the form
\begin{align}\label{eq: expansion}\nonumber
\frac{\partial \cqstate(\z,t)}{\partial t} &= \sum_{n=1}^{\infty}\frac{(-1)^n}{n!} \left(\frac{\partial^n }{\partial z_{i_1} \dots \partial z_{i_n} }\right) \left( D^{00}_{n, i_1 \dots i_n}(\z,t) \cqstate(\z,t) \right)\\ \nonumber
&  -i  [H(\z),\cqstate(\z,t)  ] + D_0^{\alpha \beta}(\z,t) \lin_{\alpha} \cqstate(\z,t) \lin_{\beta}^{\dag} - \frac{1}{2} D_0^{\alpha \beta}(\z,t) \{ \lin_{\beta}^{\dag} \lin_{\alpha}, \cqstate(\z,t) \}_+ \\ 
& + \sum_{\mu \nu \neq 00} \sum_{n=1}^{\infty}\frac{(-1)^n}{n!}  \left(\frac{\partial^n }{\partial z_{i_1} \dots \partial z_{i_n} }\right)\left( D^{\mu  \nu}_{n, i_1 \dots i_n}(\z,t) \lin_{\mu} \cqstate(\z,t) \lin_{\nu}^{\dag} \right),
\end{align}
where we define the Hermitian operator $H(\z)= \frac{i}{2}(D^{\mu 0}_0 \lin_{\mu} - D^{0 \mu}_0 \lin_{\mu}^{\dag}) $ (which  is Hermitian since $D^{\mu 0}_0 = D^{0 \mu *}_0$). Equation \eqref{eq: expansion} is what we will refer to as the \textit{Kramers-Moyal expansion} of the dynamics. We see the first line of \eqref{eq: expansion} describes purely classical dynamics, and is fully described by the moments of the identity component of the dynamics $\Lambda^{00}(\z|\z')$. The second line describes pure quantum Lindbladian evolution described by
the zeroth moments of the components $\Lambda^{\alpha 0}(\z|\z'), \Lambda^{\alpha \beta}(\z|\z')$; specifically the (block) off diagonals, $D^{\alpha 0}_0$, describe the pure Hamiltonian evolution, whilst the
components $D^{\alpha \beta}_0$ describe the dissipative part of the pure quantum evolution. Note that the Hamiltonian and Lindblad couplings can depend on the classical degrees of freedom so the second line describes action of the classical system on the quantum one. The third line contains the non-trivial classical-quantum back-reaction, where changes in the classical distribution are induced and accompanied by changes in the quantum state. We shall often write $D_n$ to describe the object with entries $D_{n,i_1 \dots i_n}^{\mu \nu}$ and we occasionally write the master equation in short-hand as
\begin{equation}
\frac{\partial \cqstate}{\partial t} = \mathcal{L}(\rho),
\end{equation}
where the \textit{superoperator} $\mathcal{L}$ is defined via the right hand side of Equation \eqref{eq: expansion}. The formal solution to the dynamics can then be written as 
\begin{equation}
\rho(t) = U(t,t_i) = \mathcal{T} \{ e^{ \int_{t_i}^t dt' \mathcal{L}(t') } \}(\rho(t_i)),
\end{equation}
where $\mathcal{T}$ denotes the time ordering operator, which is required since in the time dependent case $\mathcal{L}(t)$  operations do not commute with each other at different times. Since the dynamics is autonomous, it forms a semi-group \cite{Lindblad:1975ef}, so that $U(t,t_i) = U(t,t')U(t',t_i)$.

\section{Derivation of the path integral formalism}\label{sec: derivation}
In classical Markovian dynamics, the Kramers-Moyal expansion is used to obtain a path integral representation of the dynamics. For the reader unfamiliar with classical path integrals for open classical systems we recommend  \cite{Weber_2017} (see also \cite{risken1989fpe, Kleinert}). The path integral for quantum systems is found after Trotterizing \cite{trotter} the dynamics and inserting position and momentum resolutions of the identity -- see \cite{Sieberer_2016} for a review of quantum path integrals for open quantum systems. In the hybrid case, we shall do both simultaneously to arrive at a CQ path integral, using the short time representation of the dynamics appearing in Equation \eqref{eq: transitionAmplitude}. 

 We first derive a path integral for the most general CQ master equation to arrive at a phase space path integral, which includes an integral over response variables. The result is Equation \eqref{eq: generalAction}. In its most general form, the path integral is a complicated object, however, in section \ref{sec: derivationContinuous} we study the path integral for the class of continuous master Equations. In this case, we find that we can always integrate out the response variables to arrive at a phase space path integral alone, given by Equation \eqref{eq: continuousActionFinal}.\footnote{In section \ref{sec: configurationSpace} we discuss the sufficient conditions to derive a configuration space path integral, namely that the classical-quantum action be at most quadratic in momenta. } The resulting path integral has a natural interpretation in terms of suppressing paths away from there averaged equations of motion by an amount characterized by $D_{2}^{-1}$.\footnote{Here, and throughout, the $^{-1}$ denotes the generalized inverse of $D_2  (D_0)$, since $D_2 (D_0)$ are only required to be positive \textit{semi-definite}} Simultaneously there is decoherence the quantum system, by an amount depending on $D_0$. The decoherence diffusion trade-off \cite{dec_Vs_diff}, necessary for complete positivity of the dynamics, tells us that one cannot simultaneously make the effects of decoherence and diffusion small if there is back-reaction on the classical system.

\subsection{Derivation of phase space path integral for any CQ dynamics }\label{sec: derivationGeneral}
Let us now derive the CQ path integral for the master equation in Equation \eqref{eq: expansion}. For ease of presentation, we shall take the Lindblad operators $\lin_{\mu}$ to be functions of two canonically conjugate operators $\xq,\pq$, with $[\xq,\pq]=i $, $\lin_{\mu}(\xq,\pq)$, but the derivation also holds if they are functions of multiple operators and we can also write a coherent state path integral using similar methods to \cite{Sieberer_2016}. We use the convention that  $\langle \xq |\xq' \rangle = \delta(\xq-\xq')$ and $\langle \pq | \pq' \rangle = 2\pi \delta(\pq-\pq')$ so that $\langle \xq |\pq \rangle = e^{i \pq \xq} $.

To derive the path integral, we first Trotterize the dynamics. Defining $t_f= t_i + K \delta t$ and $t_k=k \delta t$, we use the identity
\begin{equation}\label{eq: trotterizeFull}
U(t_f,t_i) = \lim_{K \rightarrow \infty} U(t_f,t_{k-1})U(t_{k-1}, t_{k-2}) \dots U(t_2,t_i) =\lim _{K \rightarrow \infty}(1+\delta t \mathcal{L}(t_{k-1}))(1+\delta t \mathcal{L}(t_{k-2}))\dots (1+ \delta t \mathcal{L}(t_i)).
\end{equation}
In the time independent case, \eqref{eq: trotterizeFull} reduces to the familiar statement 
\begin{equation}
U(t_f,t_i) = e^{ \mathcal{L} (t_f-t_i)} = \lim_{K \to \infty} e^{ \mathcal{L} \delta t} = \lim_{K \to \infty}( 1+ \mathcal{L} \delta t)^K.
\end{equation}
We can use Equation \eqref{eq: trotterizeFull} to write the CQ state at time $t_{k+1}$ in terms of the state at time $t_k$ as 
\begin{equation}\label{eq: trotterEquation}
    \cqstate(\z_{k+1}, t_{k+1}) = \int d\z_{k } \delta( \z_{k+1}-\z_{k}) \cqstate(\z_{k},t_k) + \delta t \mathcal{L}(\z_{k+1}|\z_{k})( \cqstate(\z_{k},t_k)).
\end{equation}
Using the definition of the delta function, we can identify the right hand side of Equation \eqref{eq: trotterEquation} with that of Equation \eqref{eq: transitionAmplitude} to write 
\begin{equation}\label{eq: transitionAmplitude2}
\cqstate(\z_{k+1}, t_{k+1}) =    \frac{1}{(2 \pi)^d} \int d\u_k d\z_k \  e^{-i \u_k \cdot (\z_{k+1}-\z_k)} \left( \cqstate(\z_k,t_k)+  \sum_{n=0}^{\infty}  \delta t (i^n) \frac{ u_{k,i_1} \dots u_{k,i_n}   }{n!}  D^{\mu \nu}_{n, i_1 \dots i_n}(\z_k,t_k) \lin_{\mu}\cqstate(\z_k,t_k) \lin_{\nu}^{\dag} \right).
\end{equation}
The next step is to map the Lindblad operators acting on the CQ state to c-numbers which can be exponentiated. Just as with the quantum path integral, we first write the state in terms of the $\xq$ basis 
\begin{equation}\label{eq: stateexpansion}
    \cqstate( \z_k, t_{k}) = \int d \xql_k d \xqr_k \ \cqstate(\z_k,\xqr_k,\xql_k, t_{k}) \ | \xqr_k \rangle \langle \xql_k | ,
\end{equation}
where $ \cqstate(\z,\xqr_k,\xql_k, t_{k})= \langle \xqr| \cqstate( \z, t_{k}) | \xql \rangle $. The $+,-$ convention is such that when we calculate the expectation value of operators $\Tr{}{O^{\rind} \cqstate O^{\lind}}$, then, after using cyclicity of the trace $\Tr{}{\cqstate O^{\lind} O^{\rind} } = \Tr{}{O^{\lind} O^{\rind} \cqstate}$, the operators $O^{\lind}$ are always to the left of $O^{\rind}$. 

 Inserting \eqref{eq: transitionAmplitude2} into \eqref{eq: stateexpansion}, along with $\pqr, \pql$ resolutions of the identity (at the position $\mathbb{I}L_{\mu}\cqstate L_{\nu}^{\dag} \mathbb{I}$)  gives the following expression for the transition amplitude
\begin{align}\nonumber
\cqstate(\z_{k+1},\xqr_{k+1},\xq\lin_{k+1}, t_{k+1}) &=  \frac{1}{(2 \pi)^d}\int d \xql_k d \xqr_k  d \pql_k d \pqr_k d \u_k d\z_k e^{ \delta t \mathcal{I}( \xql_k,  \xqr_k,   \pql_k,  \pqr_k,  \u_k, \z_k,t_k) } \cqstate(\z_k,\xqr_k,\xql_k, t_{k}), 
\end{align}
where we have implicitly defined the (time-discrete) classical-quantum action
\begin{align}\nonumber
  \mathcal{I}( \xql_k,  \xqr_k,   \pql_k,  \pqr_k,  \u_k, \z_k,t_k) &=   -i \u_k \cdot \frac{(\z_{k+1}-\z_k)}{\delta t} + i\frac{(\xqr_{k+1} - \xqr)}{\delta t}\pqr - i\frac{(\xq\lin_{k+1} - \xql)}{\delta t}\pql \\
  & + \sum_{n=0}^{\infty}(i^n) \frac{ u_{k,i_1} \dots u_{k,i_n}   }{n!}  D^{\mu \nu}_{n, i_1 \dots i_n}(\z_k,t_k) \lin_{\mu}^{\rind} \lin_{\nu}^{\lind *},
\end{align}
using the shorthand $\lin_{\mu}^{\rind} := \lin_{\mu}(\xqr, \pqr) = \langle \pi^+|L_{\mu}(\phi, \pi) |\phi^+ \rangle$, and similarly for $L^{\lind}$. Taking the $t\to 0, K \to \infty$ limit, with $\delta t K = t_f-t_i$, we arrive at the path integral representation of the transition amplitude 
\begin{equation}\label{eq: pathIntegralGeneral}
\begin{split}
& \cqstate(\z, \xqr, \xql, t_f) = \\
& \lim_{K \to \infty} \int \left\{ \prod_{k=1}^K d \xq^{\pm}_k \right\}  \left\{ \prod_{k=1}^K \frac{d \pq^{\pm}_k}{2 \pi} \right\} \left\{ \prod_{k=1}^K \frac{d \u_k}{(2 \pi)^d} \right\}  \left\{ \prod_{k=1}^K d \z_k \right\}  \ e^{ \delta t \sum_{k=1}^K \mathcal{I}( \xql_k,  \xqr_k,   \pql_k,  \pqr_k,  \u_k, \z_k,t_k) } \cqstate(\z_i,\xqr_i,\xql_i, t_i),
\end{split}
\end{equation}
where it should always be understood that boundary conditions for the final state have been imposed. For ease of notation, we will write this formally as 
\begin{equation}\label{eq: pathIntegralGeneralFormal}
\cqstate(\z, \xqr, \xql, t_f) =\int  \mathcal{D} \xq^{\pm} \mathcal{D} \pq^{\pm}  \mathcal{D} \u  \mathcal{D} \z \ e^{ \mathcal{I}( \xql,  \xqr,   \pql,  \pqr,  \u, \z,t_i,t_f)}\cqstate(\z_i,\xqr_i,\xql_i, t_i) ,
\end{equation}
where 
\begin{equation}
 \mathcal{I}( \xql,  \xqr,   \pql,  \pqr,  \u, \z,t_i,t_f) =    \int_{t_i}^{t_f} dt\left[  -i \u \cdot \frac{d\z}{dt} + i\dot{\xqr}\pqr - i\dot{\xql}\pql  + \sum_{n=0}^{\infty}(i^n) \frac{ u_{i_1} \dots u_{i_n}   }{n!}  D^{\mu \nu}_{n, i_1 \dots i_n}(\z,t) \lin_{\mu}^{\rind} \lin_{\nu}^{\lind *}   \right],
\end{equation}
and the $\pm$ denotes integration over both the $\rind, \lind$ variables. Finally, we can use the normalization condition to substitute in for the coefficient $D^{00}_0$ to write the path integral in a way which reflects the structure of the master equation in \eqref{eq: expansion}, in which case we find our general expression for the \textit{CQ action}
\begin{equation}\label{eq: generalAction}
\begin{split}
    &  \mathcal{I}( \xql,  \xqr,   \pql,  \pqr,  \u, \z,t_i,t_f) =    \int_{t_i}^{t_f} dt\bigg[  -i \u \cdot \frac{d\z}{dt} + \sum_{n=1}^{\infty}(i^n) \frac{ u_{i_1} \dots u_{i_n}   }{n!}  D^{0 0}_{n, i_1 \dots i_n}(\z,t)  \\
     & +  i\dot{\xqr}\pqr - iH^{\rind}  - i\dot{\xql}\pql + iH^{\lind} + D^{\alpha \beta}_0(\z,t)  \lin_{\alpha}^{\rind}(\xqr, \pqr) \lin_{\beta}^{* \lind}  - \frac{1}{2} D^{\alpha \beta}_0(\z,t) \left( L^{* \rind}_{\beta}\lin_{\alpha}^{\rind}  + L^{* \rind}_{\beta}\lin_{\alpha}^{\rind} \right) \ \\
     & + \sum_{\mu \nu \neq 00} \sum_{n=1}^{\infty}(i^n) \frac{ u_{i_1} \dots u_{i_n}   }{n!}  D^{\mu \nu}_{n, i_1 \dots i_n}(\z,t) \lin_{\mu}^{\rind} \lin_{\nu}^{* \lind}\bigg].
     \end{split}
\end{equation}
We can break down Equation \eqref{eq: generalAction} into its familiar parts, by writing 
\begin{equation}\label{eq: cqactiondef}
 \mathcal{I}( \xql,  \xqr,   \pql,  \pqr,  \u, \z,t_i,t_f) = iS_C[\u, \z] + iS[\xqr,\pqr] - i S[\xql, \pql] + i S_{FV}[\z, \xq^{\pm}, \pq^{\pm}] + i S_{CQ}[\u, \z, \xq^{\pm}, \pq^{\pm}].
\end{equation}
In Equation \eqref{eq: cqactiondef} $S_C[\u, \z]$ is the pure classical action \cite{Weber_2017}
\begin{equation}\label{eq: genclassicalPI}
i S_C[\u, \z] =  \int_{t_i}^{t_f} dt \left[  -i \u \cdot \frac{d\z}{dt} + \sum_{n=1}^{\infty}(i^n) \frac{ u_{i_1} \dots u_{i_n}   }{n!}  D^{0 0}_{n, i_1 \dots i_n}(\z,t) \right] ,
\end{equation} 
 $S[ \xq, \pq] = \dot{\xq} \pq - H$ is a pure quantum action (written in momentum variables), which appears in the combination $iS[\xq^+, \pq^{+}] -i S[\xq^+, \pq^{+}] $ due to the bra and ket components of the density matrix. $S_{FV}[\z, \xq^{\pm}, \pq^{\pm}] $ is the Feynman-Vernon action familiar in the study of open quantum systems, describing the pure Lindbladian part of the dynamics 
\begin{equation}\label{eq: fv}
S_{FV}[\z, \xq^{\pm}, \pq^{\pm}] = - i \int_{t_i}^{t_f} dt \ D^{\alpha \beta}_0(\z,t)  \lin_{\alpha}^{\rind} \lin_{\beta}^{* \lind}  - \frac{1}{2} D^{\alpha \beta}_0(\z,t) \left( L^{* \rind}_{\beta}\lin_{\alpha}^{\rind}  + L^{* \rind}_{\beta}\lin_{\alpha}^{\rind} \right),
\end{equation}
and $S_{CQ}[\u, \z, \xq^{\pm}, \pq^{\pm}]$ is describes the novel non-trivial CQ interaction terms
\begin{equation}
iS_{CQ}[\u, \z, \xq^{\pm}, \pq^{\pm}] = \int_{t_i}^{t_f} dt \sum_{\mu \nu \neq 00} \sum_{n=1}^{\infty}(i^{n}) \frac{ u_{i_1} \dots u_{i_n}   }{n!}  D^{\mu \nu}_{n, i_1 \dots i_n}(\z,t) \lin_{\mu}^{\rind} \lin_{\nu}^{* \lind}.
\end{equation}
One sees that the quantum back-reaction on the classical system is encoded by the interaction of the Linbdlad operators with the response variables $\u$ through the coupling $ D^{\mu \nu}_{n, i_1 \dots i_n}(\z,t)$. Equation \eqref{eq: pathIntegralGeneral} is the most general path integral formulation for CQ autonomous dynamics. However, the fact that there are infinitely many terms appearing in the exponent make it potentially difficult to work with, at least exactly, and represents the fact that in general classical-quantum dynamics can involve finite sized jumps in the classical phase space. 

Nonetheless, there is an important class of dynamics for which the path integral becomes much simpler. In \cite{UCLPawula} it was proven that there are two classes of CQ dynamics in a CQ version of the Pawula theorem \cite{Pawula}. Either the CQ master equation has infinitely many moments $D_n$ or the moment series has at most two moments and describes continous dynamics in phase space.  In the case of infinitely many moments, the dynamics consists of finite sized jumps in phase space. These dynamics were introduced in \cite{blanchard1995event}, and in the case of $\z$ of finite dimension, its generality was proven by Poulin \cite{poulinPC2}. An example of a consistent continuous master equation, when the back-reaction is a constant force, first appeared in \cite{Diosi:1995qs, Diosi:2011vu}. The general form of the continuous dynamics is given by \cite{UCLPawula} 
\begin{align}\label{eq: continousME}\nonumber
\frac{\partial \cqstate(\z,t)}{\partial t} & =  \sum_{n=1}^{n=2}\frac{(-1)^n}{n!}\left(\frac{\partial^n }{\partial z_{i_1} \dots \partial z_{i_n} }\right) \left( D^{00}_{n, i_1 \dots i_n}(\z,t) \cqstate(\z,t) \right) - \frac{\partial }{\partial z_{i}} \left( D^{0\alpha}_{1, i}(\z,t) \cqstate(\z,t) \lin_{\alpha}^{\dag} \right)  - \frac{\partial }{\partial z_{i}} \left( D^{\alpha 0 }_{1, i}(\z,t) \lin_{\alpha} \cqstate(\z,t) \right)  \\ 
& -i[H(\z), \cqstate(z,t)] + + D_0^{\alpha \beta}(\z,t) \lin_{\alpha} \cqstate(\z,t) \lin_{\beta}^{\dag} - \frac{1}{2} D_0^{\alpha \beta}(\z,t) \{ \lin_{\beta}^{\dag} \lin_{\alpha}, \cqstate(\z,t) \}_+ ,
\end{align}
where completely positivity of the dynamics is equivalent to the condition that the matrix
\begin{align} \label{eq: blockMatrixPositivity}
D= 
    \begin{bmatrix}
  D_{2} & D_1\\ D^{\dag}_1 & D_0 
\end{bmatrix} 
\succeq 0
\end{align}
is positive semi-definite. It is useful to note that that since Equation \eqref{eq: blockMatrixPositivity} is in block form it follows from the Schur complement that positive semi-definiteness of $D$ is equivalent to the following conditions
\begin{equation}\label{eq: schurPositivity}
\begin{array}{l}
 D \succeq 0 \Leftrightarrow D_2 \succeq 0, D_0-D_1^{\dag} D_2^{-1} D_1 \succeq 0,\left(I-D_2 D_2^{-1}\right) D_1=0 \text { and } \\
D \succeq 0 \Leftrightarrow D_0 \succeq 0,  D_2 -D_1 D_0^{-1} D_1^{\dag} \succeq 0,\left(I-D_0 D_0^{-1}\right) D_1^{\dag}=0.
\end{array}
\end{equation}
Here $^{-1}$ denotes the generalized inverse, since $D_2, D_0$ are only required to be positive \textit{semi-definite}. For convenience, we often write the generalized inverse of an object $X$ as $X^{-1}$ but we emphasise that it should always be understood as the generalized inverse if $X$ is not invertible. As a consequence of Equation \eqref{eq: schurPositivity}, master Equations of the form of Equation \eqref{eq: continousME} must obey $  D_2 \succeq D_1 D_0^{-1} D_1^{\dag}$, which is the decoherence-diffusion trade-off explored in \citep{dec_Vs_diff}.\footnote{The factor of 2 difference with \cite{dec_Vs_diff} arises since we have not absorbed the $n!$ appearing in the master equation \eqref{eq: expansion} into the definition of the moments $D_n$.  }

For the class of continuous master equations, the action of Equation \eqref{eq: generalAction} reads  
\begin{equation}\label{eq: continuousAction}
\begin{split}
    &  \mathcal{I}( \xql,  \xqr,   \pql,  \pqr,  \u, \z,t_i,t_f) =    \int_{t_i}^{t_f} dt\bigg[  -i \u \cdot \frac{d\z}{dt} +i u_{i} D^{0 0}_{1, i} -  \frac{1}{2} u_{i}     D^{0 0}_{2, i j}u_{j}   \\
     &  i\dot{\xqr}\pqr - iH^+  - i\dot{\xql}\pql + iH^- + D^{\alpha \beta}_0  \lin_{\alpha}^{\rind} \lin_{\beta}^{* \lind } - \frac{1}{2} D^{\alpha \beta}_0 \left( L^{*\lind}_{\beta}\lin_{\alpha}^{\lind} + L^{* \rind}_{\beta}\lin_{\alpha}^{\rind} \right) \ \\
     & +  i u_{i}  D^{0 \alpha}_{1, i} \lin_{\alpha}^{* \lind}  +  i u_{i}  D^{\alpha 0 }_{1, i} \lin_{\alpha}^{\rind}\bigg],
     \end{split}
\end{equation} 
and we can identify the purely classical contribution with the \textit{Fokker-Plank} action \cite{Weber_2017, Kleinert}
\begin{equation}\label{eq: classicalContinous}
 iS_C = \int_{t_i}^{t_f} dt\left[ -i \u \cdot \frac{d\z}{dt} +i u_{i} D^{0 0}_{1, i} -  \frac{1}{2} u_{i}     D^{0 0}_{2, i j}u_{j}  \right],
\end{equation}
whilst the CQ interaction is determined via 
\begin{equation}\label{eq: CQContinous}
S_{CQ} =  \int_{t_i}^{t_f} dt \left[ u_{i}  D^{0 \alpha}_{1, i} \lin_{\alpha}^{* \lind}  +   u_{i}  D^{\alpha 0 }_{1, i}\lin_{\alpha}^{\rind} \right].
\end{equation}
Equation \eqref{eq: continuousAction} gives the general form of path integrals for any continuous autonomous CQ master equation. In the next section we show that these path integrals can be simplified further since the action is quadratic in $\u$. Therefore, we are able to integrate out the response variables $\u$ to obtain a path integral over $\xq^{\pm}, \pq^{\pm}, \z$ alone. In this case, the coupling between quantum and classical degrees of freedom can be written down directly, as opposed to being mediated by response variables, and we show this has a natural interpretation as suppressing classical paths which deviate away from their averaged path, determined by both pure classical drift and drift sourced by quantum back-reaction on the classical degrees of freedom.

To summarize this section, we have seen that by introducing a recursive short time moment expansion of the classical-quantum state, Equation \eqref{eq: transitionAmplitude2}, we are able to write down the general form of the phase space classical-quantum path integral. The general expression is given by Equation \eqref{eq: generalAction}, and for continuous dynamics we arrive at the considerably simpler path integral of Equation \eqref{eq: continuousAction}. In the general case, the action includes an integral over response variables $u$. However, for the continuous master equations the response variables couple quadratically in the action; in the next section we shall integrate these out to obtain a path integral formulation without response variables.
\section{Path integral formulation for continuous CQ master Equations without response variables}\label{sec: derivationContinuous}
In this section, we perform the integral over response variables in Equation \eqref{eq: continuousAction} to obtain a path integral representation of the continuous master equation in terms of the variables $\xq^{\pm}, \pq^{\pm}, \z$ alone. Since the pure quantum parts of the action are $\u$ independent, the relevant portion of the action, i.e, the $\u$ dependent part, is given by $S_C+ S_{CQ}$ -- as defined in Equations \eqref{eq: classicalContinous} and \eqref{eq: CQContinous}. We have to be careful since the diffusion matrix $D_2$ appearing in $S_C$ is a symmetric, positive \textit{semi-definite} matrix, as opposed to a positive definite matrix, and so the Gaussian integral over response variables is slightly less standard. 
 
 To deal with this issue, we first diagonalize the diffusion matrix by means of an orthogonal transformation $O_{ij}(\z,t)$
 \begin{equation}
     D_{2,ij}^{00}(\z,t)= \sum_k O_{ik}(\z,t)^T \lambda_k(\z,t) O_{kj}(\z,t) .
 \end{equation}
We expect the $\u$ path integral measure to be invariant under orthogonal transformations of $\u$ and so after applying the orthogonal transformation to Equation \eqref{eq: continuousAction}, or more properly to each $u$ integral in Equation \eqref{eq: pathIntegralGeneral}, we arrive at the diagonalized action for the continuous master equation
\begin{equation}\label{eq: continuousActionDiagonal}
\begin{split}
    &  \mathcal{I}( \xql,  \xqr,   \pql,  \pqr,  u, \z,t_i,t_f) =    \int_{t_i}^{t_f} dt\bigg[  -i u_i O_{ij}  \frac{dz_j}{dt} +i u_{i} O_{ij} D^{0 0}_{1, j} -  \frac{1}{2} u_{i}     \lambda_i u_{i}   \\
     &  i\dot{\xqr}\pqr - iH^+- i\dot{\xql}\pql + iH^- + D^{\alpha \beta}_0  \lin_{\alpha}^{\rind} \lin_{\beta}^{* \lind} - \frac{1}{2} D^{\alpha \beta}_0 \left( L^{* \lind}_{\beta}\lin_{\alpha}^{\lind} + L^{*\rind}_{\beta}\lin_{\alpha}^{\rind} \right) \ \\
     & +  i u_{i} O_{ij} D^{0 \alpha}_{1, j} \lin_{\alpha}^{*\lind}  +  i u_{i}  O_{ij} D^{\alpha 0 }_{1, j} \lin_{\alpha}^{\rind}\bigg].
     \end{split}
\end{equation} 
Having diagonalized the diffusion matrix $D^{00}_{2,ij}$ the $u_{i} u_{j}$ cross terms appearing in Equation \eqref{eq: continuousAction} decouple and we can perform each component $u_i$ of the $\mathcal{D} \u$ path integral separately. Since we know that $D_{2,ij}^{00}$ is a positive semi-definite matrix there are two cases to consider -- when its eigenvalues vanish $\lambda_i =0$ and when they are strictly positive $\lambda_i >0$. We split the action into terms consisting of response variables which couple to a zero eigenvalue and those which couple to a strictly positive eigenvalue 
\begin{equation}\label{eq: actionDecomposition}
\mathcal{I}( \xql,  \xqr,   \pql,  \pqr,  u, \z,t_i,t_f) = \mathcal{I}( \xql,  \xqr,   \pql,  \pqr,  u_{ \lambda=0}, \z,t_i,t_f) + \mathcal{I}( \xql,  \xqr,   \pql,  \pqr,  u_{\lambda >0}, \z,t_i,t_f).
\end{equation}
Given this decomposition, we are then able to perform the integration over the response variables in turn, starting with those associated to a zero eigenvalue.
\subsubsection{Integral over response variables $u_i$ with $\lambda_i=0$}
When $\lambda_i =0$ the action in Equation \eqref{eq: continuousActionDiagonal} is linear in $u_i$. We know that for the dynamics to be completely positive we must have \eqref{eq: schurPositivity} that 
\begin{equation}
    ( I - D_2 D_2^{-1}) D_1 =0,
\end{equation}
and as a consequence
\begin{equation}\label{eq: positiveSemiDefiniteCondition}
    ( I - \lambda \lambda^{-1} ) O D_1 =0,
\end{equation}
where $\lambda^{-1}$ is the generalized inverse of $\lambda$. We can write \eqref{eq: positiveSemiDefiniteCondition} in components as 
\begin{equation}
\sum_j ( 1 -  \delta( \lambda_i)) O_{ij}D_{1,j}=0,
\end{equation}
where $\delta(\lambda_i)$ is $0$ if $\lambda_i =0$ and one otherwise. In the case when $\lambda_i > 0$ this poses no extra restrictions. However, when $\lambda_i=0$ the $i,j$ component can only be satisfied if $\sum_j O_{ij}D_{1,j}=0$. Hence, whenever $\lambda_i=0$ the terms $u_i O_{1,j} D^{0 \alpha}_{1,j}$ - and its complex conjugate term - will not contribute in Equation \eqref{eq: continuousActionDiagonal}. This is expected: if we use a basis of classical variables such that the diffusion matrix is diagonal, then if any of the classical variables are deterministic we expect to have no quantum back-reaction on these degrees of freedom, since we know this must be of stochastic nature due to Equation \eqref{eq: schurPositivity} and the decoherence diffusion trade-off \cite{dec_Vs_diff}. 

In fact, for vanishing $\lambda_i$, the only term involving the response variable $u_i$ is given by the purely classical $S_{C}$ action, and takes the form
\begin{equation}\label{eq: SCZeroEigenvalue}
\sum_{j, i |\lambda_i =0} \left( -i u_i O_{ij} \frac{dz_j}{dt} + i u_i O_{ij} D_{1,j}^{00} \right).
\end{equation}
Performing the $u_i$ path integral over \eqref{eq: SCZeroEigenvalue} gives rise to a delta functional 
\begin{equation}\label{eq: zeroEigenValueIntegral}
\cqstate(\z, \xqr, \xql, t_f) = \int \mathcal{D} \xql  \mathcal{D} \xqr \mathcal{D} \pql  \mathcal{D} \pqr  \mathcal{D} u_{ \lambda >0}  \mathcal{D} \z \ \prod_{k| \lambda_{k} =0} \delta( O_{kj }\frac{dz_j}{dt} - O_{kj} D_{1,j} ))  \ e^{ \mathcal{I}( \xql,  \xqr,   \pql,  \pqr,  u_{ \lambda >0}, z,t_i,t_f)}\cqstate(\z_i,\xqr_i,\xql_i, t_i),
\end{equation}
where  we denote $\mathcal{D} u_{\lambda >0}$ as the remaining path integral over the variables with eigenvalue $\lambda_i >0$. $\mathcal{D} u_{\lambda >0}$ should be understood as the $K\to \infty$ limit of $\prod_{k=1}^K \prod_{i=1}^R \frac{d \vec{u}_{k,i} }{2 \pi} $, where $i=1,\dots R$ label the components of $u_{k,i}$ which have positive eigenvalue $\lambda_i>0$.

\subsubsection{Integral over response variables with $\lambda_i >0$}
Now let us consider the path integral for the response variables $u_i$ which couple to a positive eigenvalue $\lambda_i >0$. The terms in the CQ action \eqref{eq: continuousAction} which involve these response variables are given by $S_C$ and $S_{CQ}$
\begin{align}\label{eq: continuousActionDiagonalBigger}
     \sum_{i | \lambda_i \geq 0, j}\bigg[  -i u_i O_{ij}  \frac{dz_j}{dt} +i u_{i} O_{ij} D^{0 0}_{1, j} -  \frac{1}{2} u_{i}     \lambda_i u_{i}   
     +  i u_{i} O_{ij} D^{0 \alpha}_{1, j}(\z) \lin_{\alpha}^{* \lind}  +  i u_{i}  O_{ij} D^{\alpha 0 }_{1, j}(\z) \lin_{\alpha}^{\rind}\bigg].
\end{align} 
Equation \eqref{eq: continuousActionDiagonalBigger} is quadratic in $u_i$ and the quadratic terms couple to a now positive matrix $u_i\lambda_i \delta_{ij}u_i$. As such, at each timestep $t_k$ we can perform a standard Gaussian integral
\begin{equation}\label{eq: GaussianIntegral}
\int \exp d^R u \left(-\frac{1}{2} \u \cdot A \cdot \u+ i J \cdot \u \right) =\sqrt{\frac{(2 \pi)^R}{\operatorname{det} A}} \exp \left(-\frac{1}{2} J \cdot A^{-1} \cdot J\right),
\end{equation}
and integrate out the remaining response variables. Explicitly, at each time-step we can perform the integration over the remaining $u_i$ variables using Equation \eqref{eq: GaussianIntegral}, identifying $R$ as the rank of the positive definite block of $D_{2,ij}$, $A$ as the $R \times R$ matrix with elements $A_{ij} = (\delta t) \delta_{ij} \lambda_j$ and finally $J_i = O_{ij} D_{1,diff}$ where $D_{1,j}^{ diff}$ is the vector 
\begin{equation}\label{eq: driftdefinition}
  D_{1,i}^{diff}(\z, \xq^{\pm}, \pq^{\pm},t)=  D^{0 0}_{1, i}(\z,t)
     +    D^{0 \alpha}_{1, i}(\z,t) \lin_{\alpha}^*(\xql, \pql)  +    D^{\alpha 0 }_{1, i}(\z,t) \lin_{\alpha}(\xqr, \pqr)  -  \frac{dz_i}{dt}.
\end{equation}
$D_{1,j}^{ diff}$ describes the difference between the classical path $\frac{dz_i}{dt}$ and its expected drift, sourced by both quantum back-reaction $D^{0 \alpha}_{1, i}(\z) \lin_{\alpha}^*(\xql, \pql)  +    D^{\alpha 0 }_{1, i}(\z) \lin_{\alpha}(\xqr, \pqr)$, and purely classical drift $D^{0 0}_{1, i}$. 
Using Equation \eqref{eq: GaussianIntegral} we see that for each $u_i$ integration we will pick up a contribution 
\begin{equation}\label{eq: tildezintegral}
  \sqrt{\frac{ (2 \pi)^{R}}{ (\delta t)^R \operatorname{det} \lambda^+}} \exp \left(-\frac{1}{2} D_{1}^{diff} \cdot D_2^{-1} \cdot D_{1}^{diff} \delta t \right)  ,
\end{equation}
where $\operatorname{det} \lambda^+$ is defined as the product of all the positive eigenvalues of $D_{2,ij}$.

It is important to emphasise that the diffusion matrix $D_2(\z,t)$ can be $\z$ dependent, as can its eigenvalues, and so the $\lambda^+(\z,t)$ appearing in Equation \eqref{eq: tildezintegral} is also generically $\z$ dependent, describing a phase space dependent normalization factor that we shall henceforth absorb into the definition of the $\mathcal{D}z$ measure. This term is similar in nature to the additional curvature term which arises when studying classical path integrals in curved space, as well as when studying Langevin equations with multiplicative noise \cite{Dekker, Graham1977PathIF}.

All together we can integrate out all the response variables $u$, to write the classical quantum path integral in terms of the variables $\xq^{\pm}, \pq^{\pm}, \z$ alone
\begin{equation}\label{eq: continousTransitionFinal}
\cqstate(\z, \xqr, \xql, t_f) = \int \mathcal{D} \xql  \mathcal{D} \xqr \mathcal{D} \pql  \mathcal{D} \pqr   \mathcal{D} \z  \ \prod_{k| \lambda_{k} =0} \delta( O_{kj }\frac{dz_j}{dt} - O_{kj} D_{1,j} ))  \ e^{ \mathcal{I}( \xql,  \xqr,   \pql,  \pqr,  \z,t_i,t_f)} \cqstate(\z_i,\xqr_i,\xql_i, t_i),
\end{equation}
where the action takes the form 
\begin{equation}\label{eq: continuousActionFinal}
\begin{split}
    &  \mathcal{I}( \xql,  \xqr,   \pql,  \pqr,  \z,t_i,t_f) =    \int_{t_i}^{t_f} dt\bigg[  -\frac{1}{2} D_{1}^{diff} \cdot D_2^{-1} \cdot D_{1}^{diff} \\
    & + i\dot{\xqr}\pqr - iH^+  - i\dot{\xql}\pql + iH^-+ D^{\alpha \beta}_0  \lin_{\alpha}^{\rind} \lin_{\beta}^{* \lind}- \frac{1}{2} D^{\alpha \beta}_0 \left( L^{*\lind}_{\beta}\lin_{\alpha}^{\lind} + L^{*\rind}_{\beta}\lin_{\alpha}^{\rind} \right) \bigg],
    \end{split}
\end{equation} 
and we have redefined the $\mathcal{D} \z$ integration measure, which due to Equation \eqref{eq: tildezintegral}, now reads 
\begin{equation}\label{eq: measz}
    \prod_{k=1}^K \frac{d\z_k}{\sqrt{ (2 \pi \delta t)^R \operatorname{det} \lambda^+_k}}.
\end{equation}
Equation \eqref{eq: continuousActionFinal} gives the most general path integral for the entire class of continuous CQ master equations. The conditions for the underlying dynamics to be completely positive can be read off directly from Equation \eqref{eq: schurPositivity}. In particular, the underlying dynamics will be CP if and only if the Lindbladian coefficient $D_0$ is positive semi-definite $D_0 \succeq 0$, $(\mathbb{I}-D_0 D_0^{-1})D_1 =0$ and $ D_2 \succeq D_1 D_0^{-1} D_{1}^{\dag}$ where $D_0^{-1}$ is the generalized inverse $D_0$. 

The classical-quantum interaction of the action in Equation \eqref{eq: continuousActionFinal} is contained in the term 
\begin{equation}
 -\int_{t_i}^{t_f} dt \ \frac{1}{2} D_{1}^{diff}(\z, \xq^{\pm}, \pq^{\pm},t) \cdot D_2^{-1}(\z,t) \cdot D_{1}^{diff}(\z, \xq^{\pm}, \pq^{\pm},t),
\end{equation}
which, using the definition of $D_{1}^{diff}$ in Equation \eqref{eq: driftdefinition}, has a very natural interpretation as suppressing contributions to the path integral where paths $\frac{d\z}{dt}$ differ from the expected drift, $D^{0 0}_{1, i} + D^{0 \alpha}_{1, i} \lin_{\alpha}^{* \lind}+D^{\alpha 0 }_{1, i} \lin_{\alpha}^{\rind}$, which is sourced by both the pure classical evolution and the quantum back-reaction on the system. We also see that the amount one is penalized for moving away from the expected classical trajectory depends on the inverse of the diffusion matrix; if there is a large amount of diffusion, then $D_2^{-1}$ is expected to be small, and so classical paths which venture away from the expected value are penalized less. However, if there is very little diffusion then $D_{2}^{-1}$ will be large, and classical trajectories are forced to stick nearby the most likely path. Since both $D_2^{-1}$ and $D_0$ cannot simultaneously be made small by virtue of the complete positivity condition in Equation \eqref{eq: schurPositivity} there is a trade-off between the amount of diffusion in the classical system and the strength of the Lindbladian evolution of the quantum system which is characterized by the second line of Equation \eqref{eq: continuousActionFinal}. To further gain intuition for the path integral, let us now discuss some cases in which the path integral in Equation \eqref{eq: continuousActionFinal} simplifies.

\subsection{Hermitian Lindblad operators}
\label{sec: hermitian}
A particularly nice interpretation of the path integral arises when the Lindblad operators $\lin_{\alpha}$ in Equation \eqref{eq: continuousActionFinal} are Hermitian. In this case we can write the pure Lindbladian term of the path integral $S_{FV}$ in Equation \eqref{eq: fv} in terms of a negative semi-definite quadratic form
\begin{equation}
 iS_{FV} =   -\frac{1}{2}\int_{t_i}^{t_f} dt( \lin_{\alpha}^{\lind} - \lin_{\alpha}^{\rind}) D^{\alpha \beta}_0 ( \lin_{\beta}^{\lind} - \lin_{\beta}^{\rind}) ,
\end{equation}
and so Equation \eqref{eq: continuousActionFinal} reads
\begin{align}\label{eq: continuousActionHermitian}\nonumber
    &  \mathcal{I}( \xql,  \xqr,   \pql,  \pqr,  \z,t_i,t_f) =    \int_{t_i}^{t_f} dt\bigg[ i\dot{\xqr}\pqr - iH^{\rind}  - i\dot{\xql}\pql + iH^{\lind}  \\
    &   -\frac{1}{2}( \lin_{\alpha}^{\lind} - \lin_{\alpha}^{\rind}) D^{\alpha \beta}_0 ( \lin_{\beta}^{\lind} - \lin_{\beta}^{\rind})  -\frac{1}{2} D_{1}^{diff } \cdot D_2^{-1} \cdot D_{1}^{diff} \bigg].
\end{align} 
Reminding ourselves that the path integral can be used to compute the off-diagonal elements of the density operator $\cqstate(\z, \xqr, \xql, t_f)$, we see that the Lindbladian part of the action suppresses the off-diagonal elements by an amount dependent on $ \lin_{\alpha}^{\lind} - \lin_{\alpha}^{\rind}$ and the Lindbladian coupling matrix $D_0^{\alpha \beta}$. In particular when $D_0^{\alpha \beta}$ is large, we find that paths where $\lin_{\alpha}^{\lind} \sim \lin_{\alpha}^{\rind}$ are heavily preferred, causing the density matrix to decohere into the eigenbasis dictated by the Lindblad operators. On the other hand, if the magnitude of the decoherence is small, then one can maintain superpositions for a long time. However, we know from the CP conditions that long coherence times leads to necessarily large $D_2$ on the classical system, with the precise relationship determined by the strength of the quantum back-reaction $D_1$. Large diffusion means that $D_2^{-1}$ will be small, and so paths can deviate away from there average without much suppression; there is a lot of classical uncertainty. In other words, if the decoherence rate is small, so that superpositions can be maintained, then measuring the classical degree of freedom necessarily gives you little information about the coherence of the quantum state. This is precisely the reason that consistent classical-quantum dynamics exists, it is the stochastic coupling between classical and quantum degrees of freedom, and the trade-off between maintaining superpositions and classical uncertainty, which allows one to evade the no-go theorems of Feynman regarding the consistency of hybrid dynamics (see \cite{ChapelHill,Feynman:1996kb,paradoxbook,EppleyHannah} for an incomplete list of some of the original arguments on the inconsistency of classical-quantum dynamics). 

%%%%%%
\subsection{Hamiltonian drift}
\label{sec: hamdrift}
Another simplification occurs when the drift $D_1^{drift}$ is generated by a CQ Hamiltonian $H_{CQ}[\phi, \pi]= H_Q[\phi, \pi] + H_C[q,p] + V_I[q,\phi]$, as in \cite{Oppenheim:2018igd, oppenheim2020objective, Diosi2014}. We assume the interaction is minimally coupled, so that the interaction potential $V_I[q,\phi]$ depends on classical and quantum position but not momenta, whilst $H_C[q,p]$ we take to be a purely classical Hamiltonian. When the drift is generated by a Hamiltonian, the back-reaction, described by the $D_{1}^{0 \alpha}$ term in the master equation \eqref{eq: continousME}, is given by the \textit{Alexandrov-Gerasimenko bracket} \cite{Aleksandrov, Gerasimenko}
\begin{equation}\label{eq: agbracket}
    \frac{1}{2}(\{V_I, \cqstate\} - \{\cqstate, V_I \}),
\end{equation}
where for simplicity we consider only one classical degree of freedom.
Furthermore, due to the trade-off in Equation \eqref{eq: schurPositivity}, we know the full master equation must contain a term
\begin{equation}\label{eq: hamExample}
    \frac{\partial \cqstate(\q,\p)}{\partial t}  = \{H_C, \cqstate \} -i[H_Q + V_I,\cqstate]  + \frac{1}{2}\frac{\partial^2 }{\partial \p^2 }( D_{2}\cqstate)+  \frac{1}{2} \left( \{ V_{I}(\q,\xq), \cqstate \}  - \{\cqstate, V_I(\q, \xq) \} \right)  -  \frac{1}{2} D_0 \big\{ \frac{\partial V_I}{\partial q },\big\{ \frac{\partial V_I}{\partial q}, \cqstate \big\}  \big\},
\end{equation}
where $H_Q$ is a quantum Hamiltonian and complete positivity demands $4 D_2 \geq D_0^{-1}$. For the master equation in Equation 
\eqref{eq: hamExample}, the path integral action of Equation \eqref{eq: continuousActionFinal} is
\begin{align}\label{eq: continuousActionHam}\nonumber
    &  \mathcal{I}( \xql,  \xqr,   \pql,  \pqr,  \z,t_i,t_f) =    \int_{t_i}^{t_f} dt\bigg[ i\dot{\xqr}\pqr - i(H^{\rind}_Q + V_I^\rind)  - i\dot{\xql}\pql + i(H^{\lind}_Q+V_I^\lind)\\
    &   -\frac{D_0}{2} \bigg( \frac{\partial V_I}{\partial q }^{\lind} - \frac{\partial V_I}{\partial q }^{\rind} \bigg)^2 -\frac{1}{2} D_{1}^{diff } \cdot D_2^{-1} \cdot D_{1}^{diff} \bigg],
\end{align} 
where the classical-quantum interaction takes the form 
\begin{equation}\label{eq: hamDrift}
 D_{1}^{diff } \cdot D_2^{-1} \cdot D_{1}^{diff} = 
 D_2^{-1} \bigg( \{H_C, p\} + \frac{1}{2} \{V_I[q,\phi^+], p\} + \frac{1}{2}\{V_I[q,\phi^-],  p \} -\dot{p}\bigg)^2.
\end{equation}
Equation \eqref{eq: hamDrift} thus acts to suppress paths which deviate from the $\pm$ averaged Hamiltonian equations of motion. In Section \ref{sec: configurationSpace}, we show that when the action is at most quadratic in the momenta of the classical and quantum systems, one can integrate out the momentum variables to arrive at a configuration space path integral, where paths deviating from their averaged Euler-Lagrange equations are suppressed.

%%%
\subsection{When the trade-off is saturated}\label{sec: saturate}
We will now see that when one saturates the decoherence-diffusion trade-off, which we take to mean that $D_0 = D_1D_2^{-1} D_1$ in Equation \eqref{eq: schurPositivity}, a remarkable set of cancellations occur, and the path integral takes on a very simple form. In particular, it factorises, so that there is no coupling between the $\phi^+$ and $\phi^-$ fields. This reflects the fact that when the trade-off is saturated the classical-quantum dynamics keeps quantum states pure conditioned on the classical trajectories\cite{UCLHealing}. This can also be seen via the path integral approach. When the trade-off is saturated, we can expand out the classical-quantum interaction term in the path integral of Equation \eqref{eq: continuousActionFinal} to find that all of the cross terms involving $\pm$ vanish. In particular those arising from $D_0$ cancel with those arising from $D_2^{-1}$ and Equation \eqref{eq: continuousActionFinal} reduces to
\begin{equation}
  \mathcal{I}( \xql,  \xqr,   \pql,  \pqr,  \z,t_i,t_f) = \mathcal{I}_{CQ}^+( \xqr,\pqr, \z,t_i,t_f) + \mathcal{I}_{CQ}^{- *}( \xql,\pql, \z,t_i,t_f) - I_C(\z,t_i,t_f),
\end{equation}
where 
\begin{align}\label{eq: tradeoffsat}
  & \mathcal{I}_{CQ}( \phi, \pi, \z,t_i,t_f)  =\int_{t_i}^{t_f} dt \bigg[ i\dot{\xq}\pq - iH -( D_{1,i}^{00} - \frac{dz_i}{dt})(D_{2}^{-1})^{ij}D_{1,j}^{0 \alpha} \lin_{\alpha}^* - \frac{1}{2}( \lin_{\alpha} D_0^{\alpha \beta}\lin_{\beta}^* + D_{1,i}^{0 \alpha }\lin_{\alpha}^*(D_{2}^{-1})^{ij} D_{1,j}^{0 \beta} \lin_{\beta}^* ) \bigg]\\
  & I_C(\z,t_i,t_f) =\int_{t_i}^{t_f} dt \frac{1}{2}( D_{1,i}^{00} - \frac{dz_i}{dt})(D_{2}^{-1})^{ij}( D_{1,j}^{00} - \frac{dz_j}{dt}).
\end{align}
When the back-reaction is Hermitian, meaning that $D_{1,i}^{0 \alpha}\lin_{\alpha}(\phi, \pi) =\lin_{i}(\phi,\pi,z)$ is Hermitian, Equation \eqref{eq: tradeoffsat} simplifies further and after eliminating for $D_0 = D_1D_2^{-1} D_1$ its general form can be written in terms of $(D_{2}^{-1})^{ij}$ alone 
\begin{equation}\label{eq: satruateHermitian}
     \mathcal{I}_{CQ}( \phi, \pi, \z,t_i,t_f)  =\int_{t_i}^{t_f} \bigg[ i\dot{\xq}\pq - iH -( D_{1,i}^{00} - \frac{dz_i}{dt})(D_{2}^{-1})^{ij}L_j - ( L_i (D_{2}^{-1})^{ij} L_j) \bigg].
\end{equation}
An important example of dynamics where the trade-off is saturated is an ideal continuous measurement of a Hermitian operator $Z(z_t)$, where we can also allow the choice of continuous measurement, and its strength $k(z_t)$, to depend on the measurement outcome $z_t$ at time $t$. In Appendix \ref{sec: continousmeasurement} we discuss this in more detail and give a general path integral formulation for such Markovian continuous measurement and feedback procedures, extending previous work \cite{StochasticPiaction, StochasticPIcont1, PhysRevD.33.1643,PhysRevB.78.045308,PhysRevB.78.045308,PhysRevA.56.2334}.

To summarize this section, we have shown that one can integrate out the response variables for the class of continuous CQ master equations to arrive at the path integral in Equation \eqref{eq: continuousActionFinal}, which is represented in terms of the quantum phase space variables $\phi, \pi$ and the classical variables $\z$ alone. In the next section, we look at examples where we can arrive at a configuration space path integral in both quantum and classical degrees of freedom.

\section{Configuration space path integrals}\label{sec: configurationSpace}
In this section, we consider configuration space path integrals. A configuration space path integral is an important tool in understanding whether or not one can construct covariant theories (fundamental or effective) of interacting classical-quantum fields. The conditions for which we can arrive at a configuration space path integral are standard.  In ordinary quantum mechanics we are able to integrate out the quantum momenta if the path integral is at most quadratic in the quantum momentum variables. In the CQ case, if the action is quadratic in both the classical and quantum momenta then we can arrive at a full configuration space path integral. For all such dynamics, the methodology of arriving at a configuration space path integral is the same: at each time step, one completes the square in the action and performs a Gaussian path integral. In this section we focus on a class of CQ dynamics which end up having interesting configuration space path integrals. In particular, as in \cite{Oppenheim:2018igd}, we take the classical degrees of freedom to live in a phase space and we take the CQ dynamics to be generated by an interaction potential $V_I$. We shall assume the dynamics are \textit{minimally coupled}, which we take to mean that the CQ couplings depend only on position and not on momenta, and we diffuse only in momenta. This class of dynamics proves to be interesting since one finds a configuration space path integral which can be written in terms of a CQ proto-action $W_{CQ}$, as summarized by Equation \eqref{eq: continuousActionPositionSpace}.

\subsection{An explicit derivation of a configuration space path integral}
Consider a classical system $z=(\q_i,\p_j)$, coupled to a quantum system $\xq, \pq$. In this subsection we consider continuous master equations of the form 
\begin{equation}\label{eq: HamiltonianMasterEquation}
\begin{split}
    \frac{\partial \cqstate(\q,\p)}{\partial t} & = \{H_C, \cqstate \} -i[H_Q + V_I,\cqstate]  + \frac{\partial^2 }{\partial \p_i \partial  \p_j }( D_{2,ij}(
    \q_i) \cqstate)+  \frac{1}{2} \left( \{ V_{I}(\q_i,\xq), \cqstate \}  - \{\cqstate, V_I(\q_i, \xq) \} \right)  \\
    & + D_0^{\alpha \beta} (\q_i)\left( \lin_{\alpha}(\xq) \cqstate \lin_{\beta}^{\dag}(\xq) - \frac{1}{2} \{\lin_{\beta}^{\dag}(\xq) \lin_{\alpha}(\xq) , \cqstate \} \right),
\end{split}
\end{equation}
where $H_C = \sum_i \frac{\p^2_i}{2m_i} + V(\q_i)$ is the purely classical Hamiltonian, $H_Q = \frac{\pq^2}{2m_Q} + V(\xq)$ is the purely quantum Hamiltonian and $V_I(\q_i, \xq)$ is a classical-quantum interaction potential, a hybrid object which we take to only depend on the classical and quantum positions $q_i,\phi$. We also assume that $V_I$ is Hermitian, although in general this need not be the case. We take the Lindbladian coupling $D_0^{\alpha \beta} (\q_i,t)$ and the diffusion coefficient $D_{2,ij}(\q_i,t)$ to depend on the classical positions only.  We emphasize the conditions imposed on Equation \eqref{eq: HamiltonianMasterEquation} are not necessary to get a configuration space path integral; one only requires that the momentum dependence be at most quadratic. However, including $\q_i$ diffusion, or momentum dependence $\pq$ in the Lindblad operators, alters the form of the momentum integral and complicates the final form of path integral we find. 

Since we can make an arbitrary selection of Lindblad operators, we shall use this to fix $\frac{1}{2}D_{1,i}^{0 \alpha} = \delta^{i}_{\alpha}$, and take a basis  of Lindblad operators which includes $\lin_{i}= \frac{\partial V_I}{\partial q_{i}}$. With this choice, Equation \eqref{eq: HamiltonianMasterEquation} becomes 
\begin{equation}\label{eq: withLindblad}
\begin{split}
    \frac{\partial \cqstate(\q,\p)}{\partial t} & = \{H_C, \cqstate \} -i[H_Q+ V_I,\cqstate]  + \frac{1}{2}\frac{\partial^2 }{\partial \p_i \partial  \p_j }( D_{2,ij}(
    \q_i) \cqstate)+  \frac{1}{2} \left( \{ V_{I}(\q_i,\xq), \cqstate \}  - \{\cqstate, V_I(\q_i, \xq) \} \right)  \\
    & + D_0^{ij} (\q_i)\left( \frac{\partial V_I}{\partial q_{i}} \cqstate\frac{\partial V_I^{\dag}}{\partial q_{j}} - \frac{1}{2} \bigg\{\frac{\partial V_I^{\dag}}{\partial q_{j}} \frac{\partial V_I}{\partial q_{i}} , \cqstate \bigg\} \right) + \mathcal{L}_{Lindblad}( \cqstate),
\end{split}
\end{equation}
where $\mathcal{L}_{Lindblad}$ denotes the collection of pure Lindbladian terms one could include in Equation \eqref{eq: withLindblad} that are not associated to  $\frac{\partial V_I}{\partial q_{i}}$. Since these will not be accompanied by any back-reaction on the classical degrees of freedom we will ignore them, focusing on terms associated to back-reaction alone. The $\mathcal{L}_{Lindblad}$ terms could easily be added back to the final configuration space path integral by a suitable choice of Feynman-Vernon action. Since we have fixed $D_{1,i}^{0 \alpha} = \frac{1}{2}\delta^{i}_{\alpha}$ the complete positivity condition of Equation \eqref{eq: schurPositivity} is now that  $ 4 D_2 \succeq  D_0^{-1} $.

The transition amplitude for Equation \eqref{eq: withLindblad} reads 
\begin{equation}\label{eq: configspacextransition}
    \cqstate(\q,\p, \xqr, \xql, t_f) = ( \prod_{i} m_i) \int \mathcal{D} \xq^{\pm}  \mathcal{D} \pq^{\pm}   \mathcal{D}\q \mathcal{D} \p  \delta( m_i \dot{q}_i - p_i ))\ e^{ \mathcal{I}( \xql,  \xqr, \pql,\pqr, \q, \p,t_i,t_f)}\cqstate(\q,\p,\xqr_i,\xql_i, t_i),
\end{equation}
which can be read off from Equation \eqref{eq: continousTransitionFinal}. The corresponding action is given by
\begin{equation}\label{eq: continuousActionHamiltonian1}
\begin{split}
    &  \mathcal{I}( \xql,  \xqr,   \pql,  \pqr,  \q, \p ,t_i,t_f) =    \int_{t_i}^{t_f} dt\bigg[  i\dot{\xqr}\pqr - i(H_Q^{\rind}+ V_I^{{\rind}})  - i\dot{\xql}\pql + i(H_Q^{\lind} + V_I^{\lind}) \\
    & -\frac{1}{2}\bigg( \frac{ \partial V_I^{\lind}}{\partial q_{i}} - \frac{\partial V_I^{\rind}}{\partial q_{i}}\bigg) D_0^{ij} (\q,t)\bigg( \frac{ \partial V_I^{\lind}}{\partial q_{j}} - \frac{\partial V_I^{\rind}}{\partial q_{j}}\bigg) \\
    &   - \frac{1}{2} \left(\dot{\p}_i + \frac{\partial H_C}{\partial \q_i} + \frac{1}{2 } \frac{\partial V_{I}^{\lind}}{\partial \q_i} + \frac{1}{2 } \frac{\partial V_{I}^{\rind}}{\partial \q_i}\right)(D_{2}^{-1})^{ij}(\q,t)\left(\dot{\p}_j + \frac{\partial H_C}{\partial \q_j} + \frac{1}{2 } \frac{\partial V_{I}^{\lind}}{\partial \q_j} + \frac{1}{2 } \frac{\partial V_{I}^{\rind}}{\partial \q_j}\right)  
    \bigg].
\end{split}
\end{equation} 
Since there is no $\q_i$ diffusion, the pure Hamiltonian part of the classical evolution enforces the constraint $\p_i = m_i\dot{\q}_i$ via the delta functional in Equation \eqref{eq: configspacextransition}. Combined with the form of back-reaction in Equation \eqref{eq: continuousActionHamiltonian1}, we see the action suppresses paths which deviate from the $\pm$ averaged Hamilton's equations, just as in Section \ref{sec: hamdrift}. We can now perform the classical momentum integration, including over the final momenta, to get a path integral over the classical configuration space. Doing this, Equation \eqref{eq: continuousActionHamiltonian1} becomes 

\begin{equation}\label{eq: continuousActionHamiltonian}
\begin{split}
  & \mathcal{I}( \xql,  \xqr,   \pql,  \pqr,  \q ,t_i,t_f)   =  \int_{t_i}^{t_f} dt\bigg[  i\dot{\xqr}\pqr - i(H_Q^{\rind}+ V_I^{{\rind}})  - i\dot{\xql}\pql + i(H_Q^{\lind} + V_I^{\lind})  \\
    &  -\frac{1}{2}(\frac{ \partial  \Delta V_I}{ \partial q_{i}}) D_0^{ij} ( \frac{\partial \Delta V_I}{ \partial q_{j}}) - \frac{1}{2} \left(m_i \ddot{\q}_i + \frac{\partial H_C}{\partial \q_i} + \frac{\partial \bar{V_I}}{\partial q_i}\right)(D_{2}^{-1})^{ij}(\q)\left(m_j\ddot{\q}_j + \frac{\partial H_C}{\partial \q_j} +  \frac{\partial \bar{V_I}}{\partial q_j} \right)  
    \bigg].
    \end{split}
\end{equation} 
In Equation \eqref{eq: continuousActionHamiltonian} we have introduced the notation $\bar{V_I} = \frac{1}{2 }( V_I^{\lind} + V_I^{\rind})$, the $\pm$ average potential and $\Delta V_I =  V_I^{\lind} - V_I^{\rind}$ as the difference in the potential along the $\pm$ branches.

We now define a classical-quantum action $W_{CQ}$, which we refer to as the \textit{CQ proto-action}, for reasons which will become clear shortly. The proto action is defined as 
\begin{equation}\label{eq: protoAction}
    W_{CQ} =  S_C  + S_Q + S_I = \int dt L_C(q) + L_Q(\phi) + V_I(q,\phi) := \int dt \mathcal{W}_{CQ}.
\end{equation}

Where classical Lagrangian $L_C(q) = \sum_i \p_i \dot{\q}_i -H_C(\q_i, \p_i)$ and $L_Q(\phi) = \pi \dot{\phi} -H_Q(\phi, \pi)$ is the quantum Lagrangian. We then recognize the CQ interaction term of Equation \eqref{eq: continuousActionHamiltonian} as the Euler-Lagrange equations which result from varying the $\pm$ average of the proto-action $\bar{W}_{CQ}$. Specifically we can rewrite the CQ interaction in Equation \eqref{eq: continuousActionHamiltonian} in terms of variations of the $\pm$ averaged CQ proto-action $\bar{W}_{CQ}$
\begin{equation}\label{eq: interactionActionDerivation}
- \frac{1}{2} \left(m_i \ddot{\q}_i + \frac{\partial H_C}{\partial \q_i} + \frac{\partial \bar{V_I}}{\partial q_i}\right)(D_{2}^{-1})^{ij}(\q)\left(m_j\ddot{\q}_j + \frac{\partial H_C}{\partial \q_j} +  \frac{\partial \bar{V_I}}{\partial q_j} \right)  =  - \frac{1}{2} \frac{\delta }{ \delta q_i}(\bar{W}_{CQ}[q, \xq^{\pm}])(D_{2}^{-1})^{ij} \frac{\delta }{ \delta q_j}\bar{W}_{CQ}[q, \xq^{\pm}]).
\end{equation}

 In order to get a full configuration space path integral, all that remains is to do the integrals over the momentum of the quantum system. As with the standard quantum path integral, the technical requirement to be able to do this is that the action in Equation \eqref{eq: continuousActionHamiltonian} is quadratic in $\pqr$, $\pql$ so that we can perform the $\pqr$, $\pql$ integral exactly by completing standard Gaussian integrals. Since we have taken the simplest case where the only momentum dependence $\pql$, $\pqr$ comes from the Hamiltonian, the result of the momentum integration is to perform a Legendre transformation. We end up with the configuration space CQ path integral representation of the transition amplitude 
\begin{equation}\label{eq: configspaceextransitionfinal}
\cqstate(q, \xqr, \xql, t_f) = \mathcal{N} \int \mathcal{D} \xql  \mathcal{D} \xqr   \mathcal{D} \q \ e^{ \mathcal{I}( \xql,  \xqr, \q,t_i,t_f)}\cqstate(\q,\xqr_i,\xql_i, t_i),
\end{equation}
where $\mathcal{N}= \prod_i m_i$ is a normalization constant arising from the classical momentum path integral, and we have absorbed the usual factors from the Gaussian integrals into the definition of $D\phi^{\pm}$ to obtain the standard path integral measures. The action in Equation \eqref{eq: configspaceextransitionfinal} takes its final form 
\begin{equation}\label{eq: continuousActionPositionSpace}
\begin{split}
    &  \mathcal{I}( \xql,  \xqr,  \q,t_i,t_f) =    \int_{t_i}^{t_f} dt\bigg[   i\mathcal{W}_{CQ}^+  - i\mathcal{W}_{CQ}^-   -\frac{1}{2}(\frac{ \delta  \Delta W_{CQ}[q,\phi^{\pm}]}{ \delta q_{i}}) D_0^{ij} ( \frac{\delta \Delta W_{CQ}[q,\phi^{\pm}]}{ \delta q_{j}}) \\
    & - \frac{1}{2} \frac{\delta }{ \delta q_i}(  \bar{W}_{CQ}[q, \xq^{\pm}]) (D_{2}^{-1})^{ij} \frac{\delta }{ \delta q_j}(\bar{W}_{CQ}[q, \xq^{\pm}])
    \bigg].
    \end{split}
\end{equation}

Equation \eqref{eq: continuousActionPositionSpace} is remarkably simple, with everything described by the classical-quantum proto-action in Equation \eqref{eq: protoAction}. All of the CQ interaction is encoded in variations of a single classical-quantum proto-action $W_{CQ}$ and, due to the choice of Lindblad operators, the complete positivity condition is that $4 D_2 \succeq  D_0^{-1} $. The classical trajectories are suppressed away from the $\pm$ averaged equations of motion which arise from varying the proto-action by an amount depending on $D_2$, whilst there is simultaneous decoherence by an amount which depends on the difference in the equations of motion between the $\pm$ branches. In Appendix \ref{sec: peturbative} we discuss a simple toy example of the configuration space path integral and illustrate how one can use perturbative methods familiar in quantum theories to calculate CQ path integrals using Feynman diagrams.

This direct derivation of the path integral was valid for the family of master equations given by Equation \eqref{eq: HamiltonianMasterEquation}, which couple less than quadratically in the momenta. However, given the final and suggestive form of Equation \eqref{eq: continuousActionPositionSpace} it is tempting to take it as a \textit{definition} of classical-quantum dynamics and let $W_{CQ}$ be an arbitrary functional of $q, \xq^{\pm}$ \textit{and} their derivatives. Although the mapping from the path integral to the master equation may not be completed analytically, one might expect that the condition that $4 D_2 \succeq  D_0^{-1} $ is sufficient for the dynamics to be completely positive. One often does something similar in quantum theory by taking the path integral formulation to be the fundamental object of study, which often includes higher derivative terms in the action even though the mapping between master equation's and path integral can only be computed exactly when the master equation is at most quadratic in momenta. Inspired by the action in Equation \eqref{eq: continuousActionPositionSpace}, in an accompanying short paper \cite{UCLPIShort} we prove directly from path integral  methods that \textit{any} configuration space path integral of the form 
 \begin{equation}\label{eq: positiveCQ}
  \mathcal{I}( \xql,  \xqr,  \q,t_i,t_f) = \mathcal{I}_{CQ}(\q,\phi^+, t_i,t_f) + \mathcal{I}^*_{CQ}(\q,\phi^-, t_i,t_f) - \mathcal{I}_C(\q, t_i,t_f) + \int_{t_i}^{t_f} dt d \bar{x} \sum_{ \gamma} c^{\gamma}(q,t,\bar{x})( \lin_{\gamma}[\phi^+] L^{*}_{\gamma}[\phi^-])
\end{equation}
 defines completely positive CQ dynamics. In Equation \eqref{eq: positiveCQ} $c^{\gamma} >0$, $\lin_{\gamma}[\phi^{\pm}]$ can be any functional of the bra and ket variables, $\mathcal{I}_{CQ}$ determines the CQ interaction on each of the ket and bra paths and $\mathcal{I}_C(\q, t_i,t_f)$ is the purely classical action which takes real values. In order for the path integral to be convergent we impose that $\mathcal{I}_{C}$ is positive (semi) definite, as well as asking that the real part of $\mathcal{I}_{CQ}$ be negative (semi) definite. Equation \eqref{eq: continuousActionPositionSpace} is a special case of Equation \eqref{eq: positiveCQ} when $4D_0 \succeq D_2^{-1}$ is satisfied, which can be seen by expanding out the CQ action and grouping terms by $4D_0 - D_2^{-1}$ \cite{UCLPIShort}. This is true for an \textit{arbitrary} CQ proto-action $W_{CQ}$.

This result of Equation \eqref{eq: positiveCQ} is important, since it allows us to study consistent classical-quantum path integrals, even when we cannot perform the momentum integration exactly from the Hilbert space picture. Instead, using Equation \eqref{eq: positiveCQ} as a starting point, we can write down by an appropriate choice of $W_{CQ}$.

\section{Path integrals for classical fields interacting with quantum fields} \label{sec: cqfieldPI}
In this section, we comment on the path integral for classical fields interacting with quantum fields. This provides a natural arena to study the renormalization properties of classical-quantum dynamics, as well as covariant properties of classical-quantum field theories. We treat the path integral as a formal object, making no attempt to prove anything rigorously, as is often the case with field theories. 

The path integral remains largely unchanged for the case of fields. Starting with a classical-quantum master equation involving fields, one can (formally) insert various resolutions of the identity and arrive at analogous formulas for Equations  \eqref{eq: generalAction}, \eqref{eq: continuousActionFinal}, \eqref{eq: continuousActionPositionSpace}. We do not reproduce these steps here, since they are identical to those in rest of the paper. Instead we quote the final result for \textit{local}\footnote{ For dynamics which allows for spatially correlated diffusion, the most general thing one can have is to replace the  couplings like  $D_{n,i_1 \dots i_{n}}^{\mu \nu}( \z,t) \to D_{n,i_1 \dots i_{n}}^{\mu \nu}(\z,x,y,w_{i_1} \dots w_{i_n})$ and the terms appearing in the action like $D_{n,i_1 \dots i_{n}}^{\mu \nu}( \z,t)u_{i_1 } \dots u_{i_n}\lin_{\mu} \lin_{\nu}^* \to \int dx dy dw_{i_1} \dots dw_{i_n} D_{n,i_1 \dots i_{n}}^{\mu \nu}(\z, x,y, w_{i_1} \dots w_{i_n},)u_{i_1}(x_{i_1}) \dots u_{i_n} (x_{i_n})\lin_{\mu}(x) \lin_{\nu}^*(y)$. In this case, one arrives at a field theoretic version of \eqref{eq: generalAction}, but integrating out the phase space variables becomes difficult. } classical-quantum dynamics, which is to send all the couplings appearing in the action $D_{n,i_1 \dots i_{n}}^{\mu \nu}( \z,t)\to  D_{n,i_1 \dots i_{n}}^{\mu \nu}(\z,x) $, the classical variables appearing in the action $z_i \to z_i(x)$, the quantum variables to $(\xq, \pq) \to ( \xq(x), \pq(x))$ and finally we one must integrate over all space in the action $\int dt \to \int dx$, where  $x=(t,\vec{x})$. To be explicit in understanding the field theoretic case, in this section we consider a master Equation which has a Lorentz invariant path integral.\footnote{In the purely quantum case Lorentz invariant open systems, and their renormalization properties, have recently been studied in \cite{Baidya:2017eho, Avinash:2019qga}. A simple example of Lorentz invariant quantum dynamics, which was shown to be renormalizable \cite{Baidya:2017eho}, is given by the Lorentz invariant Lindblad Equation which has the action
\begin{equation}\label{eq: lorentzinvex}
\begin{split}
    &  \mathcal{I}( \xql,  \xqr,t_i,t_f) =    \int_{t_i}^{t_f} dt\bigg[   i\mathcal{L}_{Q}^+(x)  - i\mathcal{L}_{Q}^-(x) - \frac{D_0}{2}\int dx \left(\xql(x)-\xqr(x)\right)^2
    \bigg].
    \end{split}
\end{equation}} 

For a quantum field $\xq(x)$ coupled to a classical field $q(x)$, the field theoretic version of the master Equation in \eqref{eq: HamiltonianMasterEquation} is 
\begin{equation}\label{eq: fieldConfigurationSpaceME}
\begin{split}
    \frac{\partial \cqstate(\q,\p)}{\partial t} & = \{H_C, \cqstate \} -i[H_Q+ V_I,\cqstate]   +  \frac{1}{2} \left( \{ V_{I}(\q,\xq), \cqstate \}  - \{\cqstate, V_I(\q, \xq) \} \right)  + \frac{1}{2}\int d \vec{x}\ \frac{\delta^2 }{\delta \p_i(\vec{x}) \delta  \p_j(\vec{x})  }( D_{2,ij}(q, t,\vec{x}) \cqstate)\\
    & +\int  d \vec{x} \ D_0^{ij} (q, t, \vec{x})\left( \frac{\delta V_I}{\delta q_{i}(\vec{x})} \cqstate\frac{\delta V_I^{\dag}}{\delta q_{j}(\vec{x})} - \frac{1}{2} \big \{\frac{\delta V_I^{\dag}}{\delta q_{j}(\vec{x})} \frac{\delta V_I}{\delta q_{i}(\vec{x})} , \cqstate \big \} \right) ,
\end{split}
\end{equation}
where $V_I[q, \phi] = \int  d\Vec{x}\mathcal{V}_I[q,\vec{x}]$ is an interaction potential and we take the purely classical part of the dynamics to be generated by the action $S_C(q) = \int dt  \int d \vec{x} \mathcal{L}_C[q,x]$. It should be noted that Equation \eqref{eq: fieldConfigurationSpaceME} needs regularizing, since there are multiple functional derivatives acting at the same point $x$. This corresponds to the fact that in the field theoretic case the path integral will require renormalization. For the choice of dynamics in Equation \eqref{eq: fieldConfigurationSpaceME}, the path integral action is again found to be of the form in Equation \eqref{eq: continuousActionPositionSpace}
\begin{equation}\label{eq: fieldConfigurationSpaceAction}
\begin{split}
    &  \mathcal{I}( \xql,  \xqr,  \q,t_i,t_f) =    \int_{t_i}^{t_f} dt d \vec{x} \bigg[   i\mathcal{W}_{CQ}^+[q]  - i\mathcal{W}_{CQ}^-[q] 
     \\
    &     -\frac{1}{2}(\frac{ \delta  \Delta W_{CQ}}{ \delta q_{i}}) D_0^{ij}(q,t, \vec{x}) ( \frac{\delta \Delta W_{CQ}}{ \delta q_{j}}) - \frac{1}{2} \frac{\delta }{ \delta q_i}( \bar{W}_{CQ}[q, \xq^{\pm}]) (D_{2}^{-1})^{ij}(q,t, \vec{x}) \frac{\delta }{ \delta q_j}(  \bar{W}_{CQ}[q, \xq^{\pm}])
    \bigg],
    \end{split}
\end{equation}
where now $W_{CQ}[q,\phi]$ is now a space-time CQ proto-action, given by the sum of the classical, quantum, and interaction actions via Equation \eqref{eq: protoAction}.

The path integral enables us to construct CQ theories with space-time symmetries. For example, Equation \eqref{eq: fieldConfigurationSpaceAction} will describe Lorentz invariant CQ dynamics when $W_{CQ}$ is chosen to be a Lorentz invariant scalar. We study field theoretic path integrals in more detail in \cite{UCLPIShort}. There, we start with Equation \eqref{eq: positiveCQ} and prove the resulting dynamics is CP. Given the form of Equation \eqref{eq: fieldConfigurationSpaceAction}, it is natural to consider the class of dynamics 
\begin{equation}\label{eq: fieldConfigurationSpace2}
\begin{split}
    &  \mathcal{I}( \xql,  \xqr,  \q,t_i,t_f) =    \int_{t_i}^{t_f} dt d \vec{x} \bigg[   i\mathcal{W}_{CQ}^+(x)  - i\mathcal{W}_{CQ}^-(x)  
     \\
    &     -\frac{1}{2}\frac{ \delta  \Delta W_{CQ}}{ \delta q_{i}}D_0^{ij}(\q,t, \vec{x})  \frac{\delta \Delta W_{CQ}}{ \delta q_{j}} - \frac{1}{2} \frac{\delta \bar{W}_{CQ}}{ \delta q_i}(D_{2}^{-1})^{ij}(\q,t, \vec{x}) \frac{\delta \bar{W}_{CQ} }{ \delta q_j} 
    \bigg],
    \end{split}
\end{equation}
where we impose the restriction $ 4D_0 \succeq D_2^{-1}$ to ensure the action takes the form in Equation \eqref{eq: positiveCQ} and is therefore completely positive. Note, the action in Equation \eqref{eq: fieldConfigurationSpace2} takes the same form as Equation \eqref{eq: fieldConfigurationSpaceAction} but now we let $W_{CQ}$ be an \textit{arbitrary} CQ proto-action. For completeness, we include examples of Lorentz and diffeomorphism invariant dynamics which follows from Equation \eqref{eq: fieldConfigurationSpace2} in Appendix \ref{app: covariantfields}, but refer the reader to \cite{UCLPIShort} for more details.

\section{Discussion}\label{sec: discussion}
In this work we have discussed the path integral for general classical-quantum master equations, emphasising the necessary and sufficient conditions for the dynamics to be consistent and completely positive on the quantum system. In the general case we find a path integral representation of the dynamics with response variables, given by Equation \eqref{eq: generalAction}, whilst for the class of continuous master equations, we were able to integrate out the response variables to arrive at the phase space path integral of Equation \eqref{eq: continuousActionFinal}. Under certain conditions, namely when the action of Equation \eqref{eq: continuousActionFinal} is at most quadratic in classical (quantum) momenta, we can integrate out the classical (quantum) momenta to arrive at a configuration space path integral. For the case of minimally coupled Hamiltonian theories, we end up with a simple path integral representation, Equation \eqref{eq: continuousActionPositionSpace}, where the dynamics is completely encoded via the proto-action $W_{CQ}$. Given its final form, we posited that the resulting CQ action should be completely positive for an arbitrary proto-action, a result we prove via path integral methods in an accompanying short paper \cite{UCLPIShort}. We then studied the classical-quantum path integral for fields. In Appendix \ref{app: covariantfields} we give 
an explicit example of a Lorentz invariant theory, and an example of a diffeomorphism invariant theory CQ gravity based on the trace of Einstein's equation's. We now conclude by discussing possible areas for future research.

\textit{Applications of the CQ path integral.}
It would be interesting to explore possible applications of the path integral to standard quantum mechanical scenarios. Generally, we expect CQ dynamics to be a good effective description of a quantum system when one part behaves effectively classical \cite{Milburn2012}. A particularly relevant scenario is perhaps measurement based quantum control \cite{Wiseman_2001, Quantumcontrolsurvey}, or coherent quantum control with dissipative resources \cite{jacobs2014open,Milburn2012}. In Appendix \ref{sec: continousmeasurement} we introduced the path integral for the most general Markovian continuous measurement procedure one can perform. In this context, these path integral can be understood as an extension of \cite{StochasticPiaction,StochasticPIcont1} which has proved useful in simulating quantum control tasks, particularly in the strong measurement limit where saddle point approximations are valid. We also expect that the path integral could be useful for certain systems in quantum chemistry where hybrid classical-quantum coupling has previously been used to study systems beyond the mean field approximation \cite{prezhdo1999mean, kohen1998model}.

\textit{Lorentz invariant collapse models via a classical field and relativistic measurement}. We saw that the CQ path integral corresponded to a Lorentz invariant path integral which causes decoherence of the quantum state via interaction with a classical field. Using Equation \eqref{eq: positiveCQ} we are able to write down families of covariant models with a fundamentally classical field which naturally give rise to a decoherence mechanism on the quantum state. It would be interesting to explore such theories further in the context of relativistic collapse models \cite{Pearle:1988uh,PhysRevA.42.78,BassiCollapse,PhysRevD.34.470, GisinCollapse, relcollapsePearle, 2006RelCollTum}. The main difference between the CQ dynamics considered here and standard collapse models is that here, the quantum system becomes classicalised through it's interaction with a dynamical physical field, rather than an unobservable auxiliary field.

On a related note, such dynamics also corresponds to a way of constructing relativistic quantum measurement: one first identifies a classical system which acts as a measurement device and a quantum system to be measured. One then studies the effective CQ dynamics of the interacting classical and quantum systems. The details of the measurement apparatus and coupling are encoded in phenomenological parameters $D_n$, which govern the strength of the back-reaction of the quantum system and the strength of the measurement respectively.  Within this effective prescription, there would be no need to discuss when or how a measurement occurs, there is only effective dynamics of continuously interacting classical and quantum systems.  

\textit{Renormalization of classical-quantum field theories}. The renormalization of open quantum field theories was recently studied in \cite{Baidya:2017eho, Avinash:2019qga, poulinPreskill}. It was found that open $\phi^4$ theory was perturbatively renormalizable, and the complete positivity condition for the Lindblad Equation was peturbatively preserved under renormalization\cite{Baidya:2017eho}.\footnote{Though it should be said that the most general form of dynamics introduced in \cite{Baidya:2017eho} is not CP because of the negative definite Lindbladian momentum coupling $\partial_{\mu} \phi^{+} \partial^{\mu} \phi^{-}$ which enters into the master equation \cite{UCLNotpos}. Specifically, when one goes to the master Equation picture the resulting Lindbladian is not CP since the Gaussian momentum integrals are altered due to additional momentum couplings in the Lindbladian.} On the other hand, open Yukawa theory was found to be non-renormalizable \cite{Avinash:2019qga}. It would be interesting to explore the renormalization of classical-quantum field theories. This is even more highly constrained than the renormalization of open QFT, since not only does the Lindbladian coupling need to remain positive semi-definite, but so does the diffusion coupling, and for complete positivity one also demands that these be inversely related. Having a renormalizable theory of interacting classical and quantum fields would by itself be an interesting result. On the contrary, if it was found that CQ field theories are not renormalizable, and only valid as effective theories when a physical cutoff is imposed, this would have important consequences for theories which treat the gravitational field as being fundamentally classical. 

\textit{Implications for classical-quantum gravity}.
Finally, it would be interesting to explore further the consequences of the path integral in understanding theories where the gravitational field is treated classically, both as a fundamental theory, but also as an effective theory where non-Markovian effects could be incorporated. We have seen that we can construct diffeomorphism invariant theories of CQ gravity via Equation \eqref{eq:PQG-action} and it is would be worthwhile to explore these models further to better understand this type of dynamics and constraints it imposes. We were here unable to construct a complete theory giving all the components of Einstein's equations, and we leave as an open question whether or not such dynamics exists. To this end, one can consider a diffusion kernel $D_2(x,t;x't')$, or a similar decoherence kernel which couples different points in space-time.

\section*{Acknowledgements}
We would like thank Maite Arcos, Joan Camps, Isaac Layton, Emanuele Panella, Andrea Russo, Carlo Sparaciari, Barbara \v{S}oda, and Edward Witten for valuable discussions. JO is supported by an EPSRC Established Career Fellowship, and a Royal Society Wolfson Merit Award,  Z.W.D.~acknowledges financial support from EPSRC. This research was supported by the National Science Foundation under Grant No. NSF PHY11-25915 and by the Simons Foundation {\it It from Qubit} 
 Network.  Research at Perimeter Institute is supported in part by the Government of Canada through the Department of Innovation, Science and Economic Development Canada and by the Province of Ontario through the Ministry of Economic Development, Job Creation and Trade.
\bibliography{refCQ}
\bibliographystyle{apsrev}
\appendix

\section{The path integral for Markovian continuous measurement}\label{sec: continousmeasurement}
In this section we arrive at the general form of path integral representation for a continuous measurement procedure. To construct a continuous measurement, we  divide time into a sequence of intervals of length $\Delta t$, and consider a weak measurement in each interval. To obtain a continuous measurement, we make the strength of each measurement proportional to the time interval, and then take the limit in which the time intervals become infinitesimally short. We consider a continuous measurement of a Hermitian operator $Z(z_t)$, which can depend on the measurement signal $z_t$ at time $t$, and which we take to be a functional of $x,p$. The measurement signal $z_t$ is related to the measurement outcome $\alpha_t$ by $z_t = \alpha_tdt $ \cite{2006Jacob}. The measurement signal undergoes continuous evolution, whilst the measurement outcome $\alpha_t$ is wildly discontinuous, especially for weak measurements where little information is gained in each timestep. It well known \cite{wiseman_milburn_2009,2006Jacob} that for a continuous measurement the dynamics of the quantum state can be described by the set of coupled differential equations

\begin{align}\label{eq: state}
& d|\psi(t)\rangle=\left\{-k(z_t)(Z(z_t)-\langle Z(z_t) \rangle )^{2} d t+\sqrt{2 k(z_t)}(Z(z_t)-\langle Z(z_t)\rangle) d \xi \right\}|\psi(t)\rangle\\
&
\label{eq: measurement outcome1}
d z_t=\langle Z(z_t)\rangle d t+\frac{d \xi}{\sqrt{8 k(z_t)}},
\end{align}
where $z_t$ is the measurement signal, or record, and $k(z_t)$ parameterizes the measurement strength, which can in general also be $z_t$ dependent, and for simplicity we are ignoring the pure quantum evolution in comparison to the measurement dynamics. The stochastic part of the evolution, described via $d \xi_{i}$, is the standard multivariate Wiener process satisfying the Ito rules $ d\xi_i d \xi_j = \delta_{ij} dt$, $\ d \xi_i dt = 0$. The measurement outcome itself undergoes white noise dynamics $\alpha_t = \langle Z(z_t) \rangle + \frac{1}{8 k(z_t)}\frac{d \xi}{dt}$ and is discontinuous in time which is why the signal $z_t$ is preferred. From the measurement signal, one can obtain the measurement record by taking the time derivative.

From the set of non-linear stochastic differential equations given in \eqref{eq: state}, \eqref{eq: measurement outcome1}, it is possible \cite{ UCLHealing} to construct a \textit{linear} classical-quantum master equation for the combined classical-quantum state given by 

\begin{equation}\label{eq: cqmasterEqcontmeasurement}
    \frac{\partial \cqstate(z)}{\partial t} = - k(z)[Z(z),[Z(z), \cqstate(z)]] +  \frac{Z(z)}{2}  \frac{\partial \cqstate(z)}{\partial z} +   \frac{\partial \cqstate(z) }{\partial z} \frac{Z(z)}{2} + \frac{1}{16  }\frac{\partial^2}{\partial z^2}( k^{-1}(z)  \cqstate(z)) ,
\end{equation}
from which we identify $D_{1}^{0,Z} =D_{1}^{Z,0} = \frac{1}{2} $, $D_0(z) = 2k(z)$ and $D_2(z) = \frac{1}{8 k(z)}$. We can easily check that for perfect measurements, the decoherence diffusion trade-off in Equation \eqref{eq: blockMatrixPositivity} is satisfied and in fact saturated. Substituting into Equation \eqref{eq: continuousActionHermitian}, we find the corresponding action 

\begin{align}\label{eq: continuousMeasuActionApp}\nonumber
    &  \mathcal{I}( x^{\pm},  p^{\pm},  z,t_i,t_f) =    \int_{t_i}^{t_f} dt\bigg[ i\dot{x}^{+} p^{+}  - i\dot{x}^{-}p^{-}    -k(z)( Z^-(x,p) - Z^+(x,p))^2  \\
    & -4 k(z)\left(\frac{1}{2}( Z^+(x,p) + Z^-(x,p)) - \frac{dz}{dt} \right)^2 \bigg].
\end{align}    

To our knowledge, such a general form of Markovian continuous measurement path integral has not appeared in the literature, and is complementary to current approaches \cite{StochasticPiaction, StochasticPIcont1, PhysRevD.33.1643,PhysRevB.78.045308,PhysRevB.78.045308,PhysRevA.56.2334}. One could also write down a coherent state path integral for continuous measurement using the methods introduced in \cite{Sieberer_2016}, which could be of use in studying problems in optical quantum feedback \cite{PhysRevLett.70.548}. One can also allow for noisy measurements by including an appropriate Feynman-Vernon term in Equation \eqref{eq: continuousMeasuActionApp}. We should also emphasize that we have derived this from slightly different considerations than other approaches, namely by starting from complete positivity of classical-quantum dynamics. Such path integrals have proved useful in optimizations of quantum control tasks, especially in the strong measurement regime where saddle-point approximations are valid \cite{StochasticPiaction,StochasticPIcont1}.  

It is important to note that the path integral in Equation \eqref{eq: continuousActionFinal} is more general than the continuous measurement path integral of Equation \eqref{eq: continuousMeasuActionApp}. Firstly, it allows for the case where there are many measurement operators and outcomes. It also allows for noisy imperfect quantum measurements. More importantly, it allows for the case where both the classical and the quantum degrees of freedom have dynamics of their own, which is encoded in the fact that in Equation \eqref{eq: continuousActionFinal} $D_1^{diff}$ can contain purely classical evolution determined by the drift $D_{1}^{00}$, and we can also include purely Hamiltonian, and more generally Lindbladian, quantum evolution.

\section{Perturbative methods of calculating correlation functions}\label{sec: peturbative}
In this section we study a simple model of CQ interaction to illustrate how one can use standard perturbative methods to calculate classical-quantum correlation functions via CQ Feynman diagrams. 

In the main body, we considered the path integral which constructs a CQ state at a time $t_f$ from a CQ state at time $t_i$. In computing correlation functions of classical-quantum observables, the final state is not important, and so we can perform a $t_f$ integral over the classical-quantum fields to arrive at the partition function 
\begin{equation}
    Z= \int dq_f \Tr{}{\cqstate(q_f,t_f)}
\end{equation}
which for the configuration space path integral takes the form 
\begin{equation}
  Z_0=  \mathcal{N}  \int \mathcal{D} \xql  \mathcal{D} \xqr   \mathcal{D}\q \ e^{ \mathcal{I}( \xql,  \xqr, \q, t_i,t_f)}\cqstate(q_i,\xqr_i,\xql_i, t_i),
\end{equation}
where now there are no final boundary conditions imposed on the path integral. 

Formally, we can calculate correlation functions by inserting sources $J^+, J^-, J_q$ into the path integral, and taking functional derivatives with respect to the sources. The partition function of interest is therefore
\begin{equation}
    Z[J^+,J^-,J_q] = \mathcal{N}  \int \mathcal{D} \xql  \mathcal{D} \xqr   \mathcal{D}\q \ e^{ \mathcal{I}( \xql,  \xqr, \q, t_i,t_f) -i J_+ \phi^+ + iJ^- \phi^- -J_q q}\cqstate(q_i,\xqr_i,\xql_i, t_i).
\end{equation}
In general, the form of the path integral depends on the initial CQ state $\cqstate(q_i,\xqr_i,\xql_i, t_i)$ and any calculation of correlation function must be performed on a case by case basis depending on the initial state. 

However, often we are interested in stationary states, and we would like to obtain information on correlation functions over arbitrary long times by taking the limit $t_i \to -\infty, t_f \to \infty$. In open systems, as well as when calculating scattering amplitudes, it is often assumed that the initial state in the infinite past does not affect the stationary state of the system so that there is a complete loss of memory of the initial state \cite{Sieberer_2016}. Under this assumption, it is possible to ignore the  boundary term containing the initial CQ state $\cqstate(q_i,\xqr_i,\xql_i, t_i)$ and we arrive at the partition function 
\begin{equation}\label{eq: CQpartitionfunction}
   Z[J^+,J^-,J_q] = \mathcal{N}  \int \mathcal{D} \xql  \mathcal{D} \xqr   \mathcal{D}\q \ e^{ \mathcal{I}( \xql,  \xqr, \q, -\infty,\infty) -i J_+ \phi^+ + iJ^- \phi^- -J_q q}.
\end{equation}
Using equation \eqref{eq: CQpartitionfunction}, we can then use standard perturbation methods for computing correlation functions in CQ theories. 

As a simple example, consider the zero dimensional CQ theory with CQ proto-action
\begin{equation}
   W_{CQ}= - \frac{m_q^2 q^2}{2} - \frac{\lambda q^2 \phi^2}{2},
\end{equation}
and a pure quantum action given by 
$S_Q = -\frac{m_{\phi}^2\phi^2}{2}$. Assuming the decoherence diffusion trade-off is saturated, we arrive at the total action
\begin{equation}\label{eq: toyCQexact}
   \mathcal{I}[\phi^{\pm},q] = -\frac{i}{\hbar} \frac{m_{\phi}^2(\phi^{+})^2}{2} + \frac{i}{\hbar}\frac{m_{\phi}^2(\phi^{-})^2}{2} -\frac{1}{2 D_2} \left(   q^2 m_q^4 +  \frac{1}{2}\lambda^2 q^2 ((\phi^{+})^4 + (\phi^{-})^4)  + \frac{1}{2} \lambda q m_q^2 ((\phi^{+})^2 + (\phi^{-})^2)\right).
\end{equation}
We see from Equation \eqref{eq: toyCQexact} that $D_2$ in an interacting CQ theory plays exactly the same role as $\hbar$ in an interacting quantum theory. To compute correlation functions, we can therefore work peturbatively in $D_2$. Note, the double limit $D_2 \to 0, D_2^{-1}\lambda \to 0$ defines a deterministic quantum theory with no classical back-reaction.

We define the free theory as the action independent of any CQ back-reaction
\begin{equation}
   I_{free} = iS^+ - iS^-  -I_{C} =  -i \frac{m^2_{\phi}(\phi^{+})^2 }{2 \hbar} + i \frac{m^2_{\phi}(\phi^{-})^2 }{2 \hbar} -\frac{1}{2 D_2}    q^2 m_q^4.
\end{equation}
 Inserting sources, we find the partition function 
\begin{equation}
    Z_{free}[J_+,J_-,J_q] = \int d \phi^{\pm} dq e^{I_{free} - \frac{i}{\hbar}J_+ \phi^+ +\frac{i}{\hbar} J_- \phi^- -\frac{1}{D_2} J_q q },
\end{equation}
which can be performed exactly by performing each Gaussian integral individually
\begin{equation}\label{eq: allintergrals}
 Z_{free}[J_+,J_-,J_q] = (\int d \phi^{+} e^{-i\frac{m^2_{\phi}(\phi^{+})^2}{2 \hbar} - iJ_+ \phi^+
} ) (\int d \phi^{-} e^{+i\frac{m^2_{\phi}(\phi^{-})^2}{2 \hbar} + iJ_- \phi^-
} ) (\int d q e^{-\frac{1}{2 D_2}    q^2 m_q^4 - J_q q
} ).
\end{equation}
 Equation \eqref{eq: allintergrals} is evaluated as  
\begin{equation}\label{eq: freepartitionexact}
  Z_{free}[J_+,J_-,J_q] = ( \frac{-2 \pi i \hbar }{m_{\phi}^2}) e^{\frac{i J_+^2}{2 \hbar  m_{\phi}^2}}  ( \frac{2 \pi i \hbar }{m_{\phi}^2}) e^{\frac{-i J_-^2}{2 \hbar  m_{\phi}^2}}  ( \frac{\pi D_2 }{m_{q}^4}) e^{\frac{ J_q^2}{2 D_2 m_q^4 }} =Z_0 e^{\frac{i J_+^2}{2 \hbar  m_{\phi}^2}} e^{\frac{-i J_-^2}{2 \hbar  m_{\phi}^2}}  e^{\frac{ J_q^2}{2 D_2 m_q^4 }}.
\end{equation}

From Equation \eqref{eq: freepartitionexact} we can then define the propagators for the free theory
\begin{equation}
    \langle \phi^+ \phi^+ \rangle = - \frac{i  \hbar}{ m_{\phi}^2}, \  \langle \phi^- \phi^- \rangle =  \frac{i  \hbar}{ m_{\phi}^2}, \ \langle q q \rangle = \frac{D_2}{  m_{q}^4} ,
\end{equation}
and we can represent each of the propagators by the following Feynman diagrams
\begin{align}
\label{1PI2pt}
\begin{split}
\diagramone
\end{split}
\end{align}

The full partition function with the CQ interaction turned on then takes the form
\begin{equation}
  Z[J_+,J_-,J_q]  =\langle e^{\mathcal{I}_{CQ}} \rangle = \langle e^{-\frac{1}{2 D_2} \left(   \frac{1}{2}\lambda^2 q^2 ((\phi^{+})^4 + (\phi^{-})^4)  + \frac{1}{2} \lambda q m_q^2 ((\phi^{+})^2 + (\phi^{-})^2)\right)} \rangle 
\end{equation} and we can perform an asymptotic expansion of the CQ interaction in terms of $D_2$ to arrive at the usual Feynman rules for computing correlation functions. Specifically, for terms in the action like $ \lambda_{nml} \phi^{n}_+ \phi^{m}_- q^l$, we assign the vertex with value $\lambda_{nml} n! m! l!$ to each topologically distinct diagram.

As an example, the CQ interaction term $q (\phi^{\pm})^2$ in Equation \eqref{eq: toyCQexact} has two tri-verticies with strength $-\frac{2!}{4 D_2} \lambda m_q^2 $ and can be represented by the diagrams
\begin{align}
\label{1PI2pt}
\begin{split}
\interactiondiagram
\end{split}
\end{align}
We also have the sextic $q^2 (\phi^{\pm})^4$ interaction with vertex value $-\frac{\lambda^2 4! 2! }{4D_2 }$ which is assigned to each of the following diagrams
\begin{align}
\label{1PI2pt}
\begin{split}
& \interactiondiagramsix \\
& \interactiondiagramsixdash .
\end{split}
\end{align}

\section{Lorentz and diffeomorphism invariant CQ path integrals}\label{app: covariantfields}
For completeness, in this section we give examples of Lorentz and diffeomorphism invariant CQ path integrals which can be constructed using the result of \cite{UCLPIShort}. Namely that the path integral with action in Equation \eqref{eq: fieldConfigurationSpace2}
\begin{equation}\label{eq: appfieldConfigurationSpace2}
\begin{split}
    &  \mathcal{I}( \xql,  \xqr,  \q,t_i,t_f) =    \int_{t_i}^{t_f} dt d \vec{x} \bigg[   i\mathcal{W}_{CQ}^+(x)  - i\mathcal{W}_{CQ}^-(x)  \\
    &     -\frac{1}{2}\frac{ \delta  \Delta W_{CQ}}{ \delta q_{i}}D_0^{ij}(\q,t, \vec{x})  \frac{\delta \Delta W_{CQ}}{ \delta q_{j}} - \frac{1}{2} \frac{\delta \bar{W}_{CQ}}{ \delta q_i}(D_{2}^{-1})^{ij}(\q,t, \vec{x}) \frac{\delta \bar{W}_{CQ} }{ \delta q_j} 
    \bigg],
    \end{split}
\end{equation}
is completely positive for an \textit{arbitrary} CQ proto-action $W_{CQ}$ when the restriction $ 4D_0 \succeq D_2^{-1}$ is imposed.

\subsection{Lorentz invariant classical-quantum theories}

Lorentz invariant path integrals for open quantum systems have appeared in \cite{poulinPreskill,Baidya:2017eho}. The fact that Lorentz invariant CQ dynamics exists can already be seen in Equation \eqref{eq: fieldConfigurationSpaceAction} by taking $D_2$ to be a constant and picks a Lorentz invariant interaction $V_I$.
As another example, in Equation \eqref{eq: appfieldConfigurationSpace2} we could pick a proto-action based on the stress energy tensor of the quantum matter $T_{\mu \nu}$. For example, we can consider a proto-action
\begin{equation}
    W_{CQ}[q,\phi] = \int d^4 x \mathcal{L}_Q[\phi] - \frac{1}{2}\partial_{\mu}q \partial^{\mu} q - \lambda \eta^{\mu \nu}T_{\mu \nu}[\phi](x) q(x)
\end{equation}
 where $\mathcal{L}_Q$ is a purely quantum Lagrangian and we have taken the classical action for $q$ to be a mass-less Klein-Gordon scalar field. We find Lorentz invariant dynamics which causes decoherence of the quantum state according to
\begin{equation}\label{eq: LorentzInvariantActionApp}
\begin{split}
     \mathcal{I}( \xql,  \xqr,  \q,t_i,t_f) & =    \int_{t_i}^{t_f} dt d \vec{x} \bigg[   i\mathcal{L}_{Q}^+(x)  - i\mathcal{L}_{Q}^-(x)     -\frac{\lambda^2 D_0[q]}{2}(T^-(x) - T^+(x))^2 \\
     & - \frac{1}{2D_2[q]} \left( -\partial q_{\mu}\partial^{\mu} q(x) + \frac{\lambda}{2}(T^+(x) + T^-(x))\right)^2
    \bigg],
    \end{split}
\end{equation}
where $T=\eta^{\mu \nu}T_{\mu \nu}$ is the trace of the stress energy tensor. Such dynamics essentially amounts to a Lorentz invariant collapse model, where in Equation \eqref{eq: LorentzInvariantActionApp} the collapse occurs dynamically due to interaction with a classical field. In particular, when the proto-action is based on the stress energy tensor there is an amplification mechanism by which states with small energy maintain coherence, whilst macroscopic objects decohere. The key difference between the CQ dynamics considered here and standard collapse models is that here the quantum system classicalizes due to its interaction with a dynamical classical field. It turns out that we can further arrive at diffeomorphism invariant CQ dynamics by taking the CQ interaction potential $W_{CQ}$ to be related to a gravitational action, which we now show.

\subsection{Diffeomorphism invariant theories of classical-quantum gravity}\label{sec: gravityfield}
 
Let us now explore some of the consequences for classical-quantum theories of gravity and construct an example of a diffeomorphism invariant dynamics. The natural thing to do is to attempt to construct covariant classical-quantum dynamics which approximates Einstein dynamics. Since the paths away from $\frac{\delta }{ \delta q_i}(  \bar{W}_{CQ}[q, \xq^{\pm}])$ are exponentially suppressed by an amount depending on $D_2^{-1}$, the most likely path will be those for which 
\begin{equation}
\frac{\delta }{ \delta q_i}( \bar{W}_{CQ}[q, \xq^{\pm}]) \approx 0.
\end{equation}
As such, in order to approximate Einstein gravity, we can take $W_{CQ}$ to be the sum of the Einstein Hilbert action $S_{EH}[g]= \frac{1}{16 \pi} \int dx \ \sqrt{-g} R[g]$  and a matter action $S_m[g,\phi]$. For example,  in the case of gravity coupled a scalar-field, we take the proto-action to be 
\begin{equation}
    W_{CQ}[q,\phi]= \frac{1}{16 \pi} \int dx\ \sqrt{-g} R[g] + \int \mathrm{d} x \sqrt{-g} \left[-\frac{1}{2} g^{\mu \nu} \nabla_{\mu} \phi \nabla_{\nu} \phi-\frac{1}{2} m^{2} \phi^{2}\right],
\end{equation}
 the sum of the Einstein Hilbert and Klein-Gordon actions.

When the proto-action is related to the gravitational action $W_{CQ}[g,\phi] = S_{EH}[g] + S_m[g,\phi]$ we have
\begin{equation}\label{eq:KGaction}
\frac{\delta }{ \delta g_{\mu \nu}}( \bar{W}_{CQ}[g, \xq^{\pm}])  = - \frac{\sqrt{-g}}{16 \pi}( G^{\mu \nu} - \frac{1}{2}( 8 \pi (T^{ \mu \nu })^{\rind} + 8 \pi  (T^{ \mu \nu })^{\lind}  ),
\end{equation}
and so paths would be exponentially suppressed away from the $\pm$ averaged Einstein equation by an amount depending on $D_2$. 

Taking the full CQ action of Equation \eqref{eq: fieldConfigurationSpace2} gives 
\begin{equation}\label{eq:PQG-action}
\begin{split}
    &  \mathcal{I}[ \xql,  \xqr,  g_{\mu \nu}] =    \int dx \bigg[   i\mathcal{L}_{KG}^{\rind}  - i\mathcal{L}_{KG}^{\lind}     -\frac{\det(-g)}{8}( T^{\mu \nu + } - T^{\mu \nu - }) D_{0,\mu \nu\rho \sigma}(T^{\rho \sigma+ } - T^{\rho \sigma- }) \\
    & - \frac{\det(-g)}{512 \pi^2} ( G^{\mu \nu} - \frac{1}{2}( 8 \pi (T^{ \mu \nu })^{\rind} + 8 \pi  (T^{ \mu \nu })^{\lind}  ) D_{2, \mu \nu \rho \sigma}^{-1}[g] ( G^{\rho \sigma} - \frac{1}{2}( 8 \pi (T^{ \rho \sigma })^{\rind} + 8 \pi  (T^{ \rho \sigma })^{\lind}  )
    \bigg].
    \end{split}
\end{equation}
The interaction is fully characterized by the tensor densities $D_{0,\mu \nu\rho \sigma}$ and $D_{2,\mu \nu \rho \sigma}^{-1}$ which must be satisfy $4D_0 \succeq D_2^{-1}$. With the present notation, this means $v^{\mu \nu} (4D_{0,\mu \nu\rho \sigma} - D_{2,\mu \nu \rho \sigma}^{-1}) v^{\rho \sigma} \geq 0$ for any matrix $v^{\rho \sigma}$. Constructing CQ theories which approximates Einstein dynamics then corresponds to choosing a suitable $D_0$, $D_2$, which satisfy the complete positivity condition and gives rise to classical paths which tend to stick towards Einstein's equations.

The simplest choice which is completely positive is to take $D_{0,\mu \nu\rho \sigma} = D_0 g^{-1/2} g_{\mu \nu} g_{\rho \sigma}$, in which case one finds a diffeomorphism invariant CQ theory of gravity in which paths deviating from the trace of Einstein's equations are suppressed, along with a simultaneous decoherence according to the trace of the stress energy tensor.  In the Newtonian limit, where the trace of the stress energy tensor is dominated by its mass term, this acts to decohere the quantum state into mass eigenstates. This is a related amplification mechanism to that used in spontaneous collapse models \cite{Pearle:1988uh,PhysRevA.42.78,BassiCollapse,PhysRevD.34.470, GisinCollapse, relcollapsePearle, 2006RelCollTum}, but here the decoherence mechanism arises as a consequence of treating the gravitational field classically and imposing diffeomorphism invariance on the CQ action.

Because of this decoherence and diffusion, the path integral defined by Equation \eqref{eq:PQG-action} does not suffer from the same pathologies as the standard semi-classical Einstein's equation's $
    G_{\mu \nu} = 8 \pi G \langle T_{\mu \nu} \rangle$ \cite{Oppenheim:2018igd, UCLHealing}. The reason the standard semi-classical equations fail is that they do properly account for correlations between the classical and quantum degrees of freedom. For example, the semi-classical equations predict the same thing for a test particle falling in a gravitational field produced by a planet in superposition of approximately orthogonal states as it does for a gravitational field produced by a planet in a statistical mixture of the same states \cite{GisinCollapse, Oppenheim:2018igd}.

On the other hand, the action in \eqref{eq:PQG-action} includes the correlation between the matter and gravitational degrees of freedom via the CQ interaction term on the second line --
though its not a complete theory since it is based on the trace of Einstein's equations. For example, consider starting in an initial state describing a planet in superposition of left and right $|L\rangle, |R \rangle$ states, then the action of the decoherence term will be to enforce (in the Newtonian limit) that the quantum state decoheres into mass eigenstates -- meaning that after the decoherence time the planet will be found on either the left, or the right. Because of the CQ interaction, paths where the quantum state decoheres into being on the left are correlated with the classical paths in which the gravitational field is sourced by a planet on the left $T^{\mu \mu}_L$, and similarly for paths which decohere to the planet being found on the right. This healthier type of semi-classical dynamics was explored in detail in \cite{UCLHealing} and it would be interesting to explore this further in the context of a diffeomoprhism invariant theory.

The challenge in constructing a complete theory is to obtain the transverse parts of the Einstein equation, whilst still ensuring the path integral over classical metrics remains negative definite so that the path integral converges. In Lorentzian signature, this doesn't appear to be possible within the current framework, since it amounts to constructing a positive definite metric out of the metric tensor alone. One could instead choose a $D_{0,\mu \nu\rho \sigma}$ which is non purely geometric, in which case it either introduces a preferred background, or must be made dynamical. It is also possible to consider a $D_{0,\mu \nu\rho \sigma}(x,x')$ which is a positive definite kernel in space-time coordinates $x,x'$ in which case one has stochastic processes which are correlated in space time, and the CQ interaction terms take the form

\begin{equation}\label{eq: decoNonlocal}
  -\frac{1}{2}\int dx dx' \ \frac{\delta \Delta W_{CQ}}{\delta g_{\mu \nu}(x)}D_{0,\mu \nu\rho, \sigma} (x,x') \frac{\delta \Delta W_{CQ}}{\delta g_{\rho \sigma}(x')} -\frac{1}{2}\int dx dx' \ \frac{\delta \bar{W}_{CQ}}{\delta g_{\mu \nu}(x)}D^{-1}_{2,\mu \nu\rho \sigma}(x,x')  \frac{\delta \bar{W}_{CQ}}{\delta g_{\rho \sigma}(x')} .
   \end{equation}

One could alternatively look to study dynamics which retains some space-time symmetry, such as spatial diffeomorphism invariance, but breaks full diffeomorphism invariance. This could be useful when studying effective dynamics where full diffeomorphism invariance may be broken. For example, using the ADM decomposition of the metric $g_{\mu \nu} =-N^2dt^2 + h^{ab}( N^a dt + dx^a)( N^b dt + dx^b)$, then by making use of the lapse and shift vectors and the family of positive definite de-Witt metrics $G_{a b c d}[\beta]=\frac{\sqrt{h}}{2}\left(h^{a c} h^{b d}+h^{a d} h^{b c}-2 \beta h^{a b} h^{c d}\right)$, $\beta < 1/3$, one could attempt to write down CQ actions invariant under foliation preserving diffeomorphisms, just as one does in Horava gravity \cite{Horava:2009uw}.

As either an effective or fundamental theory, it is important to understand the subtleties of constraints in CQ theories of gravity. The phase space path integral is usually where one sees the constraint structure. For example in GR the phase space path integral linear in the lapse $N$ and the shift $N^i$ variables, and performing the functional integral over them leads to a delta functional which enforces the constraints.  However, by inspecting the gravitational action in Equation \eqref{eq:PQG-action} one sees that the usual constraint structure will be necessarily altered: the CQ evolution is quadratic in $W_{CQ}$ and hence for gravity will be quartic in momenta. As a result, going to the momentum picture is more involved and we leave the study of constraints as an interesting direction for future work. It would also be interesting to study CQ constraints in simpler theories with gauge invariance, such as those with reparameterization invariance, but this is also beyond the scope of the current work.

As an effective theory, one could also look to study non-Markovian extensions of CQ dynamics. If the dynamics is non-Markovian but time-local, then one expects a large class of master equation to take a similar form to that found in \cite{Oppenheim:2018igd}, i.e, Equation \eqref{eq: expansion}, but the diffusion and decoherence coupling need not be positive semi-definite \cite{Hall2014, Breuer2016}. This could lead to Equation \eqref{eq:PQG-action} retaining full diffeomorphism invariance in the non-Markovian regime. 
\end{document}